\documentclass[%
 reprint,
 amsmath,amssymb,
 aps,prd,nofootinbib,showpacs
]{revtex4-1}

\usepackage{graphicx}
\usepackage{dcolumn}
\usepackage{bm}
\usepackage{color}
\usepackage{amsmath}
\usepackage{amssymb}
\usepackage{fullpage}
\usepackage{hyperref}
\newcommand{\aap}{Astron. Astrophys.}
\newcommand{\mnras}{Mon. Not. R. Astron. Soc.}
\newcommand{\lcdm}{$\Lambda$CDM }

\newcommand{\mpl}{m_{\mbox{\tiny{Pl}}}}
\newcommand{\Beq}{\begin{equation}\begin{aligned}}
\newcommand{\Eeq}{\end{aligned}\end{equation}}

\newcommand{\vp}{\varphi}

\newcommand{\bk}{{\bf{k}}}

\newcommand{\bq}{{\bf{q}}}

\newcommand{\mH}{{\mathcal{H}}}
\newcommand{\vvp}{{\vec{\varphi}}}
\newcommand{\mS}{{\mathcal{S}}}
\newcommand{\lsim}{\mathrel{\hbox{\rlap{\lower.55ex\hbox{$\sim$}} \kern-.3em \raise.4ex \hbox{$<$}}}}
\newcommand{\gsim}{\mathrel{\hbox{\rlap{\lower.55ex\hbox{$\sim$}} \kern-.3em \raise.4ex \hbox{$>$}}}}

\begin{document}

\title{Probing early-universe phase transitions with CMB spectral distortions}

\author{Mustafa A.~Amin$^1$}\email{mamin@ast.cam.ac.uk}
\author{Daniel Grin$^{2}$,}
\affiliation{$^1$Kavli Institute for Cosmology at Cambridge, Madingley Rd, Cambridge CB3 OHA, United Kingdom}
\affiliation{$^{2}$Department of Astronomy \& Astrophysics, University of Chicago, Chicago, Illinois 60637, U.S.A}

\date{\today}
\begin{abstract}
Global, symmetry-breaking phase transitions in the early universe can generate {\it scaling} seed networks which lead to metric perturbations. The acoustic waves in the photon-baryon plasma sourced by these metric perturbations, when Silk damped, generate spectral distortions of the cosmic microwave background (CMB). In this work, the  chemical potential distortion ($\mu$) due to scaling seed networks is computed and the accompanying Compton $y$-type distortion is estimated. The specific model of choice is the $O(N)$ nonlinear $\sigma$-model for $N\gg 1$, but the results remain the same order of magnitude for other scaling seeds. If CMB anisotropy constraints to the $O(N)$ model are saturated, the resulting chemical potential distortion $\mu \lesssim 2\times 10^{-9}$.

 \end{abstract}

\maketitle
\section{Introduction}
The primordial plasma likely underwent symmetry-breaking phase transitions. Some, like electroweak symmetry breaking, are nearly certain to have occurred \cite{Filipe:1994pc,Durrer:2001cg}. Others, like the Peccei-Quinn \cite{Peccei:1977hh} and supersymmetric phase transitions, are less established, but may be related to solutions of fine tuning problems, reheating after inflation \cite{Allahverdi:2010xz} and the physics of dark matter \cite{Jungman:1995df}.

If the broken phase is degenerate, causally disconnected regions (`Hubble patches') end up in different vacua forming topological defects and ``non-topological textures" through the Kibble mechanism. Vacua with dimension $n=0,1,2,$ or $3$ form domain walls, cosmic strings, monopoles, and textures respectively (Refs.  \cite{Albrecht:1992sb,Durrer:2001cg} and references therein). If $n\ge 4$, non-topological textures form \cite{Spergel:1990ee}: these still have significant gradient energy. These ``seeds" create long-range gravitational potential wells and thus source density perturbations in matter and radiation.

Long after the phase transition, scaling sets in: the statistics of this \textit{scaling seed} network always look the same relative to the only scale in the problem, the current physical horizon \cite{Durrer:2001cg}. More precisely, the power spectra of all seed-induced metric perturbations  are self-similar, obeying $k^{3}P_{\Phi}(k)\propto  f^2(k\eta)$ for some analytic function $f(k\eta)$, where $k$ is the wave number and $\eta$ is the conformal time \cite{Hu:1996yt}. 

Seeds excite scalar, vector, and tensor metric fluctuations \cite{Durrer:2001cg,Pen:1993nx,Turok:1997gj}. These are sourced before and after horizon crossing and so seeds are \textit{active sources} of cosmological fluctuations. The resulting perturbations in matter and the baryon-photon plasma are similar to isocurvature fluctuations \cite{Hu:1996vq}. 

Cosmic microwave background (CMB) power spectra sourced only by \textit{scaling seeds} would be out of phase and less coherent than those generated by adiabatic perturbations \cite{Durrer:2001cg, Perivolaropoulos:1992gy,Durrer:1993db,Pen:1994pe,Durrer:1993db,Crittenden:1995xf,Albrecht:1995bg,Magueijo:1995xj,Hu:1996vq,Turok:1996ud,Magueijo:1996px,Durrer:1997ep,Seljak:1997ii,Durrer:1997rh,Albrecht:1997nt,Durrer:1998rw,Fenu:2013tea,Albrecht:1992sb,Jeong:2004ut,Pogosian:2006hg}. CMB anisotropy and large-scale structure (LSS) data \cite{Durrer:2001cg,Durrer:1995sf,Albrecht:1997mz} thus limit the fractional contribution of seeds to the primordial fluctuation power to be $\lsim 0.01-0.05$ \cite{Hu:1996yt,Durrer:1997te,Sakellariadou:1999ru,Bevis:2007gh,Ade:2013xla} on large scales, which are dominated by the standard adiabatic (and likely inflationary) power spectrum with index $n_{s}\simeq 0.96$.

\begin{figure*}[t] 
   \centering
   \includegraphics[width=5in]{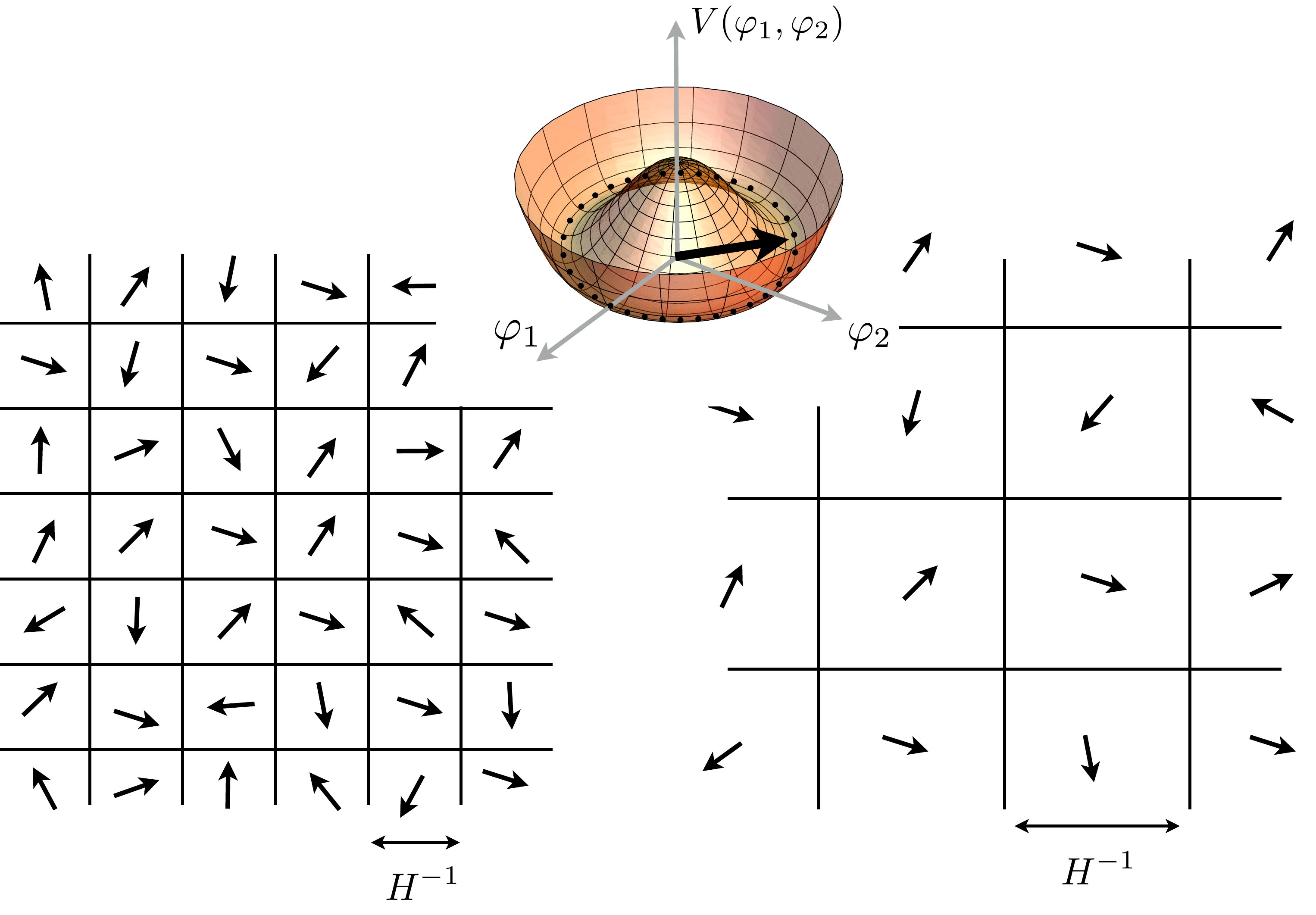} 
   \caption{After the phase transition the scalar fields reside on the vacuum manifold, picking out an (uncorrelated) ``angle" in each Hubble patch. As the universe expands, the local gradients in the field vanish on subhorizon scales.  This process leads to scaling network of the scalar field, that looks the same when compared to the contemporary Hubble horizon \cite{marcdraw}.}
   \label{fig:Sombrero}
\end{figure*}

Smaller scales with wavenumber $k\gsim50\,{\rm Mpc}^{-1}$ are beyond the reach of CMB anisotropy and existing LSS measurements. Fortunately, distortions of the CMB frequency spectrum (\textit{spectral distortions}) away from a perfect blackbody are an interesting probe of fluctuations on these scales. They provide a possible window on inflationary and scaling seed-sourced contributions on small scales. In general, acoustic waves damp by diffusion \cite{Silk:1967kq,1991ApJ...371...14D}. The energy lost from acoustic motion and injected at redshifts $z\lsim 2 \times 10^{6}$ cannot be perfectly thermalized \cite{Zeldovich:1969ff,1970Ap&SS...7...20S,1982A&A...107...39D,1986ApJ...300....1S,1991A&A...246...49B}, imprinting spectral distortions on the CMB \cite{Zeldovich:1969ff,1970Ap&SS...7...20S}. These spectral distortions allow the recovery of some of the information lost from the anisotropy spectrum. 

Modes ($50\,{\rm Mpc}^{-1}\lsim k\lsim 10^{4}\,{\rm Mpc}^{-1}$) that damp when Compton scattering is efficient ($5\times 10^{4}\lsim z\lsim 2\times 10^{6}$) will generate chemical potential ($\mu$) distortions. Modes with $k\lsim 50~{\rm Mpc}^{-1}$ will damp later and generate Compton $y$-type distortions, likely to be buried under a $y$-distortion
 from reionization \cite{Hu:1993tc,Hu:1993gc}. Distortions of the $\mu$-type, however, are a robust probe of primordial physics. 

The nearly perfect CMB blackbody measured by COBE FIRAS imposed the limits $\mu \leq 9\times 10^{-5}$ and $y\leq 1.5 \times 10^{-5}$ \cite{Mather:1993ij}, putting to rest hints of large spectral distortions and  early structure formation \cite{1988ApJ...329..567M,1991ApJ...367..420B,1989ApJ...344...24A,1991MNRAS.248...52B}. Progress in experimental techniques, described in the PIXIE and PRISM satellite proposals  \cite{Kogut:2011xw,2013arXiv1306.2259P}, could allow detection of spectral distortions $3-4$ orders of magnitude smaller than the FIRAS limits. The damping of acoustic modes sourced by standard adiabatic fluctuations with $n_s\simeq 0.96$ and no running generates spectral distortions of the CMB. The signal is roughly $\mu\simeq 1\times 10^{-8}$ and $y \simeq 2 \times 10^{-9}$ \cite{Chluba:2012gq}, providing a potentially attractive target for PIXIE/PRISM.  As discussed below, scaling seeds also lead to spectral distortions, and could be detected by these missions.

In this paper, we calculate the spectral distortion imprint of acoustic waves generated by scaling seeds. Our model of choice is the $O(N)$ nonlinear $\sigma$-model. In this model, a global symmetry $O(N)$ of a scalar field multiplet $\vec{\varphi}(\vec{x},\eta)$ is broken in a phase transiton, with the field then restricted to the vacuum manifold with an expectation value of $v$ (see Fig. \ref{fig:Sombrero}). This model offers a computational advantage: the evolution of the scaling seed network has a closed-form solution in the large-$N$ limit, known to be reasonably accurate from simulations \cite{Spergel:1990ee,Turok:1990gw,Turok:1991qq,Jaffe:1993tt,2010PhRvD..81l3504F}. 

After the phase transition, fluctuations in the \textit{direction} of the vacuum state (Goldstone modes) are always being ironed out for wavelengths within the horizon, as shown schematically in Fig. \ref{fig:Sombrero} \cite{marcdraw}. Nevertheless, the network always has fluctuations with fixed variance on the horizon scale, and is thus invariant when lengths are scaled with the cosmic expansion. Such fluctuations (isocurvature in nature) also generate temperature anisotropies in the CMB. Limits to the nonlinear $\sigma$-model obtained from recent \textit{Planck} satellite measurements\footnote{Somewhat more stringent limits could be imposed using a combination of \textit{Planck} and BICEP2 data \cite{Durrer:2014raa}, if the $B$-mode polarization anisotropy detected by BICEP2 is confirmed to be primordial \cite{Ade:2014xna}.} of CMB anisotropies \cite{Ade:2013xla} can be expressed as $v^{2}/(\sqrt{N}m_{\rm pl}^{2})\lsim 1.3\times 10^{-5}$. 

We perform a detailed calculation of the gravitational potential induced by scaling fields, and then self-consistently determine the response of photons, baryons, neutrinos, and cold dark matter (CDM), following modes through horizon crossing on to acoustic oscillation and diffusion damping.  For the $O(N)$ model, we find that 
\begin{equation}
\mu \simeq 12 \times \frac{1}{N}\left(\frac{v}{\mpl}\right)^{\!4},
\end{equation} and estimate (up to decoupling)
\begin{equation}
y\simeq 2.4 \times \frac{1}{N}\left(\frac{v}{\mpl}\right)^{\!4}.
\end{equation}

Saturating the anisotropy constraint on the $O(N)$ models, the resulting $\mu$ and $y$-type spectral distortion signals are
\begin{eqnarray}
\mu &\lesssim& 2\times 10^{-9},\\
 y&\lesssim &4\times 10^{-10}.
 \end{eqnarray}
A quantity like $v^{4}/(Nm_{\rm pl}^{4})$ normalizes other active source models, and so we expect our prediction to apply up to a factor of order unity to all such models, as argued later. 

Our signature is smaller than the standard adiabatic case by an order of magnitude, but comparable to the distortion generated by adiabatic cooling of electrons \cite{Chluba:2012gq}. There is no direct evidence, however, that the standard scenario holds at spectral distortion scales; if the adiabatic power spectrum dies off at high $k$, a phase-transition-generated spectral distortion signal could dominate. This may even be the case in a wide class of single-field inflationary models, due to running of the power spectrum on small scales \cite{Clesse:2014pna}.\footnote{The tension between {\it{Planck}} and the BICEP2 results (if confirmed) can be ameliorated by a strong negative running of the spectral index \cite{Ade:2014xna}, reducing the overall $\mu$-distortion from inflationary perturbations \cite{Chluba:2012we}.}  Moreover, when more precise measurements of spectral distortions are made, the detailed shape of the distortion could potentially disentangle different early-universe scenarios \cite{Chluba:2013pya} that lead to observable spectral distortions.

In addition to a potential signal from scaling seeds, spectral distortions could reveal the shape of the primordial power spectrum \cite{Hu:1994bz,Chluba:2012we}, the proportions of adiabatic/isocurvature modes \cite{Dent:2012ne,2013MNRAS.434.1619C}, or the presence of primordial magnetic fields \cite{Jedamzik:1999bm,Miyamoto:2013oua,Kunze:2013uja,Ganc:2014wia}, all on much smaller scales than current measurements. Spectral distortions generated by cosmic strings are estimated in Ref. \cite{Tashiro:2012pp}. Spectral distortions are also sensitive to other processes at $z\lsim 10^{6}$, like dark matter annihilation/decay \cite{McDonald:2000bk,Zavala:2009mi,Chluba:2011hw} or early star formation \cite{1994ApJ...436..456H,Hu:1993tc,Hu:1993tc}. Missions like PIXIE and PRISM could open a window to measuring $\mu$, $y$ in detail and also detecting recombination-era line emission \cite{2006MNRAS.371.1939R,2008A&A...485..377R,AliHaimoud:2012da,Sunyaev:2009qn}, motivating recent work on thermalization during this era \cite{Chluba:2004cn,Stebbins:2007ve,Chluba:2008dg,2008A&A...488..861C,Chluba:2011hw,Chluba:2012gq,Khatri:2012rt,Pajer:2012dw,Chluba:2013wsa}.

Our  plan for the rest of the paper is as follows. We begin in Sec.~\ref{sec:nonlinearsigma} by developing the nonlinear $\sigma$-model in the large-$N$ limit, including a computation of the seed metric power spectra. In Sec.~\ref{sec:specd}, we lay out perturbation evolution equations for photons, baryons, neutrinos, and CDM, and then compute the damped evolution of acoustic modes sourced by large-$N$ scaling seeds. We then compute the resulting $\mu$ and $y$ distortions. In Sec.~\ref{sec:discussion} we generalize the $O(N)$ model and provide estimates of the spectral distortion in a broader class of models. We conclude in Sec.~\ref{sec:conclusions}. We discuss technical issues of seed-correlator coherence  in the Appendix.

\section{Nonlinear $\sigma$-model}
\label{sec:nonlinearsigma}
We consider $N$ real scalar fields $\varphi_a(\mathbf{x},\eta)$ which are governed by the following Lagrangian after a global phase transition \cite{Durrer:2001cg}:
\Beq
\label{eq:Lagrangian}
\mathcal{L}=-\frac{1}{2}\partial_\mu\vec{\varphi}\cdot\partial^\mu\vec{\varphi}-\lambda(\vec{\varphi}\cdot\vec{\varphi}-v^2)^2.
\Eeq
The $N$-component field vector $\vec{\varphi}=\left\{\varphi_{1},...,\varphi_{N}\right\}$ settles into different vacua (`directions')  in causally disconnected Hubble patches. 

After the phase transition, the fields are then accurately modeled by assuming that they are on the vacuum manifold everywhere.  For $N\ge 4$ the bulk of field energy is contained in field gradients along the vacuum direction \cite{Spergel:1990ee,Turok:1991qq,Pen:1993nx,Durrer:2001cg}. Using the above Lagrangian with the constraint $\vec{\varphi}\cdot\vec{\varphi}=v^2$, the equations of motion (EOMs) in Cartesian field-space are
\begin{equation}
\Box \vec{\varphi}+\frac{\partial^{\mu}\vec{\varphi}\cdot \partial_{\mu}\vec{\varphi}}{v^{2}}\vec{\varphi}=0.
\end{equation} This is the well-known nonlinear $\sigma$-model.

Note that although $\vec{\varphi}\cdot\vec{\varphi}=v^2$ after the transition, there are gradients in $\vec{\varphi}$ from one Hubble patch to another. Since the transition is taken to occur after inflation, these gradients subsist and source gravitational fields that influence the motion of photons, baryons, neutrinos, and cold dark matter. We wish to compute the homogeneous and perturbed evolution of these scalar fields, as well as the gravitational potentials induced by them.

Before providing a detailed calculation we first estimate the gravitational potentials generated by the seeds. After the phase transition, the field energy is dominated by gradients on Horizon scales. The energy density due to $\varphi_i$, a single-field component  with wavenumber $k\sim aH$, is $a^{-2}(\nabla \varphi_{i})^{2}\propto H^2  \varphi_{i}^{2}$. Summing over $N$ field components, the variance of the energy-density fluctuation $(\delta \rho_{\varphi})^{2}\sim N H^4  \varphi_{i}^{4}$. Since the scalar-field multiplet is restricted to be on the vacuum manifold: $\sum_{i=1}^N \varphi_i^2=v^2$, we can estimate the variance of a single field component as $ \varphi_{i}^{2}\sim v^{2}/N$. Hence the variance of the density fluctuations becomes $(\delta \rho_{\varphi})^{2}\sim H^4v^4/N$. With the density perturbation in hand, we can use the Poisson equation $(k/a)^2\Phi_S\sim \mpl^{-2}\delta\rho_\varphi$ to estimate the seed-generated gravitational potential. On horizon scales ($k\sim aH$), this yields 
\Beq
\Phi_S\sim v^{2}/(\sqrt{N} m_{\rm pl}^{2}).
\Eeq 

We now turn to a more careful treatment of the field evolution and the gravitational potential generated by them.

\subsection{Evolution of $N$ scalar fields}

We use a flat FRW metric with conformal time $\eta$ (i.e. $d\eta=dt/a$) and cosmological scale factor $a(\eta)$. The EOMs, ignoring metric fluctuations, are then given by \cite{Spergel:1990ee}
\Beq \label{eq:eomfield}
\ddot{\vec{\varphi}}+\frac{\alpha}{\eta}\dot{\vec{\varphi}}-\nabla^{2}\vec{\varphi}=\frac{\partial_{\mu}\vec{\varphi}\cdot\partial^{\mu}{\vec{\varphi}}}{v^{2}}\vec{\varphi}\equiv T(\eta,\mathbf{x})\vec{\varphi},
\Eeq where $\alpha=2 d\ln{a}/d\ln{\eta}$. During the epoch of interest, it is accurate to treat the universe as a mixture of matter and radiation, and so $a(\eta)$ is given by\footnote{Note that this scale factor is $\textit{not}$ normalized to $a=1$ today, and so care must be taken when converting present-day best-fit cosmic densities to their early-time values.}
 \begin{equation}
a(\eta)=\frac{\eta}{\eta_{\rm eq}}+\frac{1}{4}\left(\frac{\eta}{\eta_{\rm eq}}\right)^{2}.
\end{equation}
Hence $\alpha=2$ for radiation domination and $\alpha=4$ for matter domination. Since seeds are assumed to form a small fraction of the total energy density, the metric perturbations induced by scaling seeds are small perturbations of the background geometry. It is thus safe to use the FRW equation of motion for the scalar field; this is the ``stiff approximation" \cite{Veeraraghavan:1990yd}.

The EOM, Eq.~(\ref{eq:eomfield}), is cubic, and may be simulated numerically, but to simplify our treatment, we make the scaling \textit{ansatz} that the trace of the scalar stress-energy tensor $T(\eta,\mathbf{x})$ is replaced by a spatially averaged quantity $\overline{T}$ which scales appropriately with $\eta$, the only dimensionful quantity in the problem \cite{Spergel:1990ee,Turok:1991qq,Pen:1993nx}:
\begin{equation}
\overline{T}(\mathbf{x},\eta)=\frac{T_{0}}{\eta^{2}}.\end{equation} The intuition behind this \textit{ansatz} is that there is only one physical scale in the problem, the horizon $\eta$, and thus any product of first-order time derivatives must scale accordingly, with some normalization. Similarly, the average picks out $k\sim \eta^{-1}$ for spatial gradients. We assume that this does not undermine our assumption that the field is everywhere on the vacuum manifold. 

We will later show that this is sufficient to guarantee that the perturbations scale in the sense discussed in the introduction. This hypothesis turns out to be quite accurate in capturing both the homogeneous and perturbed stress-energy tensor of the scaling seeds \cite{Spergel:1990ee}. 

Using Eq.~(\ref{eq:eomfield}), we now obtain the scalar-field evolution, evolving forward from initial field amplitudes $\vec{\varphi}_{\mathbf{k}}(\eta_{t})$ at the conformal time $\eta_{t}$ of the phase transition \cite{Spergel:1990ee,Turok:1991qq}:
\begin{eqnarray}
\vec{\varphi}_{\mathbf{k}}&=&\phi_{k}(\eta)\vec{\varphi}_{\mathbf{k}}(\eta_{t}),\label{eq:modevolve}\\
\phi_{k}(\eta)&=&\left(\frac{\eta}{\eta_{t}}\right)^{\left(1-\alpha\right)/2}\frac{J_{\nu}(k\eta)}{J_{\nu}(k\eta_{t})},\label{eq:modefunc}\\
\nu^{2}&=&T_{0}+\frac{1}{4}\left(\alpha-1\right)^{2}.\label{eq:besselindex}
\end{eqnarray}
The mode function $\phi_k(\eta)$ describes the time evolution of the field for a comoving wave vector $k$, while $\vec{\varphi}_{\mathbf{k}}(\eta_{t})$ captures its stochastic nature at the moment of transition. By definition, $\phi_k(\eta_{t})=1$. We have ignored a decaying mode here.\footnote{Our condition that the fields are restricted to the vacuum manifold everywhere might be violated near the phase transition. As a result the above solution is strictly valid only at times $\eta\gg \eta_t$. Details near the phase transition do not affect the behavior of the fields on the scales we are interested in at late times, for which $k\ll \eta_t^{-1}$.}

We also assume that at $\eta_t$, $\varphi_a(\mathbf{x},\eta_t)$ is correlated on subhorizon scales but uncorrelated on superhorizon scales. In Fourier space, this behavior is equivalent to a white noise power spectrum on superhorizon scales and a rapidly decaying spectrum on subhorizon scales. Explicitly we assume
\Beq
\left \langle \varphi_{\mathbf{k}}^{a}(\eta_{t})\varphi_{\mathbf{q}}^{b*}(\eta_{t})\right \rangle=&\left(2\pi\right)^{3}\delta^{ab}\delta(\mathbf{k}-\mathbf{q})P_{\varphi}(k,\eta_t)\\
P_{\varphi}(k,\eta_{t})=&\left\{ \begin{array}{ll}A_{\alpha}\neq 0 &\mbox{if $k\eta_{t}\leq 1$},\\0&\mbox{if $k\eta_{t}\geq 1$}.\end{array}  \right.\label{eq:earlyspec}
\Eeq
Small changes in shape, smoothness etc. of this assumed power spectrum do not affect our answers significantly. We normalize the power spectrum by imposing the condition that the field is on the vacuum manifold at all times,
$v^{2}=\left \langle |\vec{\varphi}(\mathbf{x},\eta)|^{2}\right \rangle$, yielding \cite{Spergel:1990ee}:
\Beq
v^{2}=NA_{\alpha}\left(\frac{\eta}{\eta_{t}}\right)^{-(2+\alpha)}\int_{0}^{\eta/\eta_{t}} \frac{d^{3}x}{\left(2\pi\right)^{3}} \frac{J_{\nu}^{2}(x)}{J_{\nu}^{2}(x\eta_{t}/\eta)}.\nonumber
\Eeq
We are interested in epochs long past the phase transition, and so $\eta\gg \eta_{t}$. Taking this limit in the above integral, time independence of the l.h.s forces us to set  
\Beq
\nu=1+\alpha/2,
\Eeq
whereas for the r.h.s to equal $v^2$, we have to set
\begin{equation}
\label{eq:Aalpha}
A_\alpha=
\begin{cases} 
3.63\times\frac{1}{N}v^2\eta_t^3 & \quad \alpha=2,
\\ \\
2.38\times\frac{1}{N}v^2\eta_t^3& \quad  \alpha=4.
\end{cases}
\end{equation}
Furthermore, using the relationship between $\nu$ and $\alpha$ in Eq.~(\ref{eq:besselindex}) we get $T_{0}=(3/4)(1+2\alpha)$. We can now determine the time-dependent field power-spectrum using Eqs.~(\ref{eq:modevolve})-(\ref{eq:Aalpha}):
\begin{eqnarray}
P_{\varphi}(k,\eta)=\left(\frac{\eta}{\eta_t}\right)^{\!\!\!1-\alpha}\!\!\left[ \frac{J_{1+\alpha/2}(k\eta)}{J_{1+\alpha/2}(k\eta_{t})}  \right]^{\!2}\!\!P_{\varphi}(k,\eta_{t}).
\end{eqnarray}

\begin{figure}[t]\centering
\includegraphics[width=2.5in]{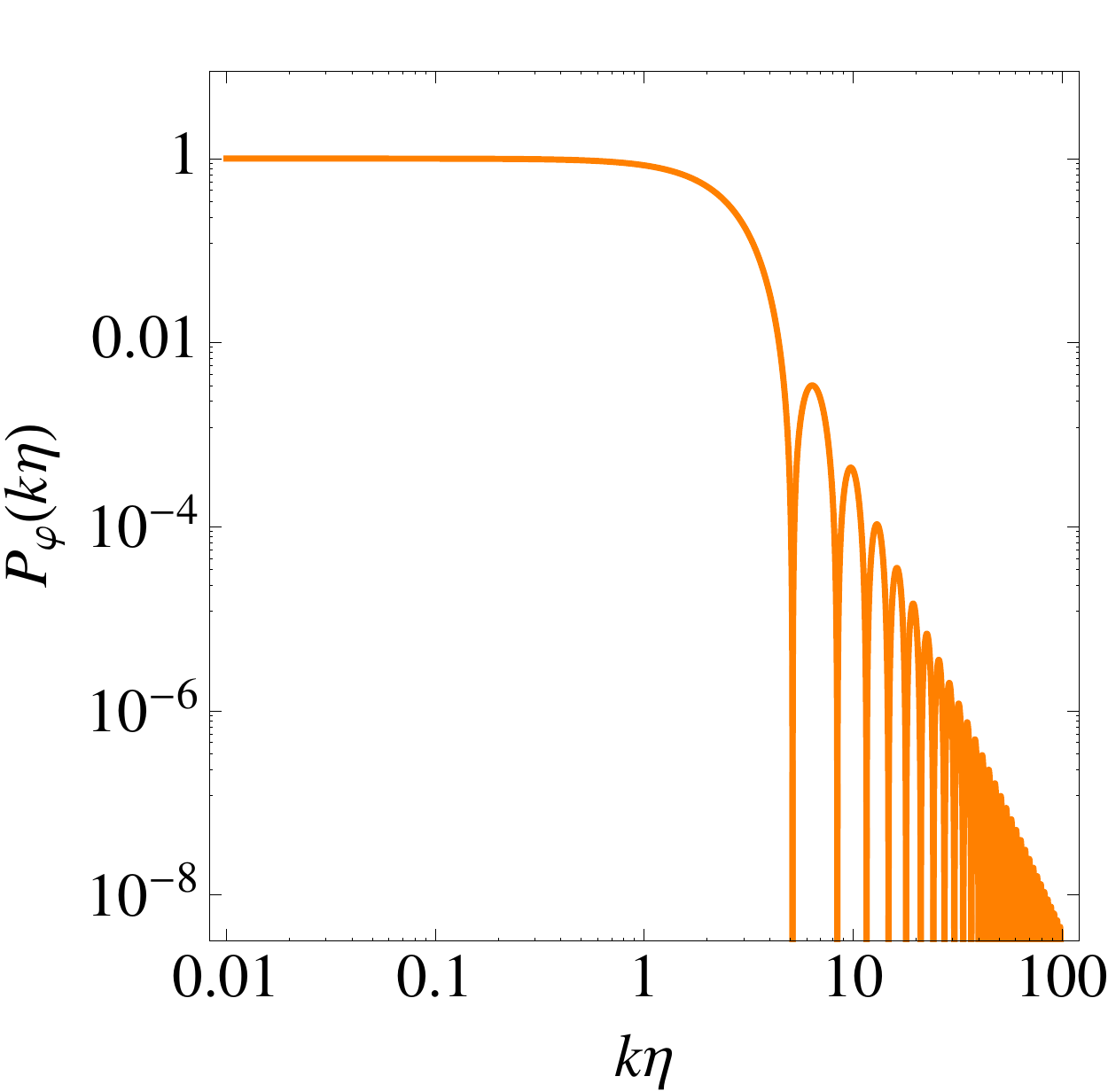}
\caption{$O(N)$ scalar-field power spectrum in a radiation-dominated universe (arbitrary normalization). The power spectrum depends only on $k\eta$, and is thus self-similar. There exists an additional cutoff (not apparent in this plot) at $k\eta=\eta/\eta_t\gg 10^2$, set by the initial time of the phase transition $\eta_t$.}
\label{fig:FieldSpectraRD}
\end{figure}

The power spectrum is computed and shown in Fig. \ref{fig:FieldSpectraRD}. The results depend only on $k\eta$ and are thus self-similar, with a cutoff at $k\eta\gtrsim1$ reflecting the erasure of perturbations through vacuum realignment as different regions come into causal contact \cite{Spergel:1990ee}.  This network thus exhibits the scaling phenomenon discussed in the Introduction. 

\subsection{Metric perturbations}
To calculate the evolution of acoustic waves in the baryon-photon fluid sourced by the scaling seeds, we must compute the gravitational potential generated by the scaling seeds. We work with the metric in conformal Newtonian gauge
\Beq
ds^{2}=a^{2}[-(1+2\Phi)d\eta^{2}+(1-2\Psi)d\mathbf{x}\cdot d\mathbf{x}],\Eeq where we have neglected vector and tensor perturbations for simplicity's sake. In Fourier space, the Einstein equations are then\begin{eqnarray}
\delta G^{0}_{0}&=&~6\frac{\mH^{2}}{a^{2}}\left[\Phi_{\mathbf{k}}+\frac{\dot{\Psi}_{\mathbf{k}}}{\mH}+\frac{k^{2}}{3\mH^{2}}\Psi_{\mathbf{k}}\right]=\frac{\delta T^{0}_{0}}{m_{\rm Pl}^{2}},\label{eq:eina}\\
\delta G^{0}_{j}&=&-2i\frac{\mH}{a^{2}}k_{j}\left[ \Phi_{\mathbf{k}}+\frac{\dot{\Psi}_{\mathbf{k}}}{\mH} \right]=\frac{\delta T^{0}_{j}}{m_{\rm Pl}^{2}},\label{eq:einb}\\
\delta G^{i}_{j}&=&\frac{1}{a^{2}}k_{i}k_{j}(\Phi_{\mathbf{k}}-\Psi_{\mathbf{k}})\nonumber\\&+& 2\frac{\mH^{2}}{a^{2}}\left [ \frac{\ddot{\Psi}_{\mathbf{k}}}{\mH^{2}}-\frac{k^{2}}{2\mH^{2}}(\Phi_{\mathbf{k}}-\Psi_{\mathbf{k}})+\frac{1}{\mH}\left(\dot{\Phi}_{\mathbf{k}}\right.\right.\nonumber\\ &+&\left.\left.2\dot{\Psi}_{\mathbf{k}}\right)-\Phi_{\mathbf{k}}\left(1-\frac{2\ddot{a}}{\mH^{2}}\right)\right ]\delta_{j}^{i}=\frac{\delta T_{j}^{i}}{m_{\rm Pl}^{2}},\label{eq:einc}
\end{eqnarray} where the stress-energy tensor here includes contributions from seeds, baryons, photons, neutrinos, and dark matter. Linearly combining Eqs.~(\ref{eq:eina})-(\ref{eq:einb}) and applying the anisotropic stress projection operator $\hat{k}_{i}\hat{k}_{j}-\frac{1}{3}\delta_{i}^{j}$ to Eq.~(\ref{eq:einc}), we obtain
\begin{eqnarray}
\Psi_{\mathbf{k}}=\frac{1}{2m_{\rm pl}^{2}}\frac{a^{2}}{k^{2}}\left( \delta T^{0}_{0}-3i\frac{\mH}{k} \hat{k}_{j}\delta T^{0}_{j}\right),\\ \Phi_{\mathbf{k}}=\Psi_{\mathbf{k}}+\frac{3}{2m_{\rm pl}^{2}}\frac{a^{2}}{k^{2}}\left(\hat{k}_{i}\hat{k}_{j}-\frac{1}{3}\delta_{i}^{j}\right)\delta T_{j}^{i}.
\end{eqnarray}
Repeated indices are summed over.
We now obtain the seed potentials in a fixed realization of $\varphi(\vec{x},\eta)$. 

It is helpful to decompose the stress-energy tensor into seed and non-seed components, that is $T_{\mu}^{\nu}=\sum_{n}{}^{\left(n\right)}T_{\nu}^{\mu}+S^{\mu}_{\nu}$, where $n$ denotes baryons, cold dark matter, neutrinos, or photons. This allows the seed component of the metric perturbation to be separately evaluated. The total metric perturbation can then be computed after allowing matter and radiation components to respond to the seed potentials. In terms of the seed stress tensor (which is denoted $\delta S^{\mu}_{\nu}$ as we assume that it has no homogeneous background value), we then have the seed induced potentials
\begin{align}
\Psi_{\mathbf{k},\rm S}=&\frac{1}{2m_{\rm pl}^{2}}\frac{a^{2}}{k^{2}}\left( \delta S^{0}_{0}-3i\frac{\mH}{k} \hat{k}_{j}\delta S^{0}_{j}\right),\label{eq:Psi}\\ 
\Phi_{\mathbf{k},\rm S}=&\Psi_{\mathbf{k},\rm S}+\frac{3}{2m_{\rm pl}^{2}}\frac{a^{2}}{k^{2}}\left(\hat{k}_{i}\hat{k}_{j}-\frac{1}{3}\delta_{i}^{j}\right)\delta S_{j}^{i}.
\label{eq:Phi}
\end{align}
The seed stress tensor can be calculated from Eq. (\ref{eq:Lagrangian}) with $\vec{\varphi}\cdot\vec{\varphi}=v^2$ as follows:
\begin{align}
S_{0}^{0}=&-\frac{1}{2a^{2}}\left[\dot{\vec{\varphi}}\cdot\dot{\vec{\varphi}}+\nabla\vec{\varphi}\cdot\nabla\vec{\varphi}\right]=-\delta \rho,\\
S_{i}^{0}=&-\frac{1}{a^{2}}\dot{\vec{\varphi}}\cdot \partial_{i}\vec{\varphi},\\
S_{j}^{i}=&\frac{1}{a^2} \partial_{i}\vec{\varphi}\cdot \partial_{j}\vec{\varphi}+\frac{1}{2a^{2}}\delta^{i}_{j}\left[ \dot{\vec{\varphi}}\cdot\dot{\vec{\varphi}}-\nabla\vec{\varphi}\cdot\nabla\vec{\varphi}\right].
\label{eq:SeedStess}
\end{align}

Using Eqs. (\ref{eq:Psi})-(\ref{eq:SeedStess}), expressions for the scalar gravitational  potentials in Fourier space are:
\begin{align}
\Psi_{\bk,{\rm S}}
=&-\frac{1}{4\mpl^2k^{2}}\int \frac{d^3q}{(2\pi)^3}\left[\dot{\vvp}_{\bq+\bk}\cdot\dot{\vvp}_\bq^*\right. \label{eq:psiconvo}\\+&\left.\bq\cdot(\bq+\bk)\vvp_{\bq+\bk}\cdot\vvp_\bq^*-6\frac{\mH}{k}(\hat{\bk}\cdot\bq)\dot{\vvp}_{\bq+\bk}\vvp_\bq^*\right],\nonumber\\
\Phi_{\bk,\rm S}=&-\frac{1}{4\mpl^2k^2}\int \frac{d^3q}{(2\pi)^3}\left[\dot{\vvp}_{\bq+\bk}\cdot\dot{\vvp}_\bq^*\right.\nonumber \\+&\left.3\left(q^2-\bk\cdot\bq-2(\hat{\bk}\cdot\bq)^2\right)\vvp_{\bq+\bk}\cdot\vvp_\bq^*\right.\nonumber \\-&\left.6\frac{\mH}{k}(\hat{\bk}\cdot\bq)\dot{\vvp}_{\bq+\bk}\vvp_\bq^*\right].\label{eq:phiconvo}
\end{align}

\begin{figure}[t] 
   \centering
   \includegraphics[width=2.8in]{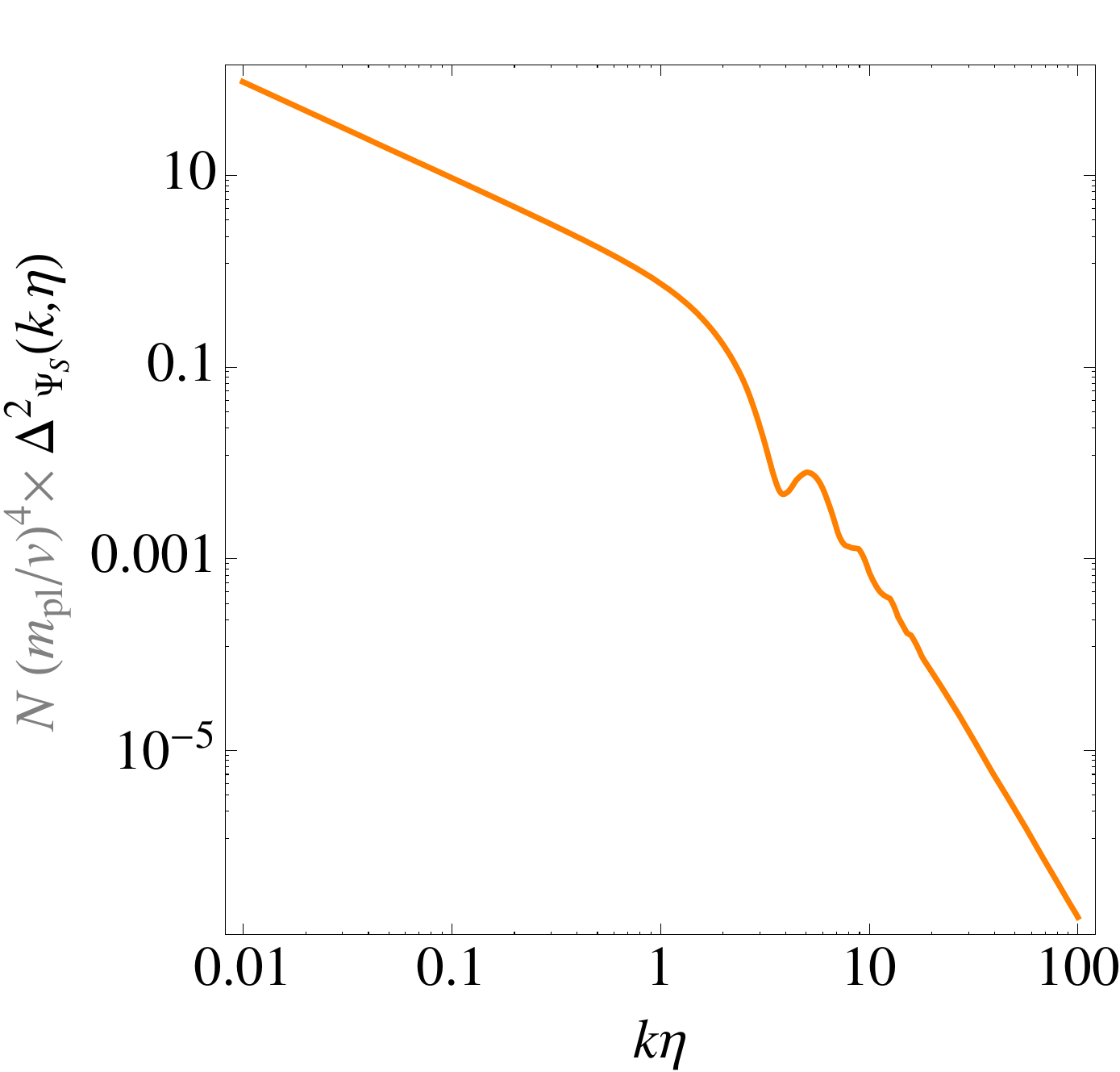} 
   \caption{The dimensionless curvature potential spectrum due to the scalar field only in a radiation-dominated universe (under the scaling ansatz for the solutions). The spectrum on super horizon scales is $\approx 0.94(k\eta)^{-1}$ whereas on subhorizon scales it is $\approx 4.2(k\eta)^{-4}\ln \left[0.56 (k\eta)\right]$.}
   \label{fig:PotentialSpectra1RD}
\end{figure}

\begin{figure}[t] 
   \centering
   \includegraphics[width=2.8in]{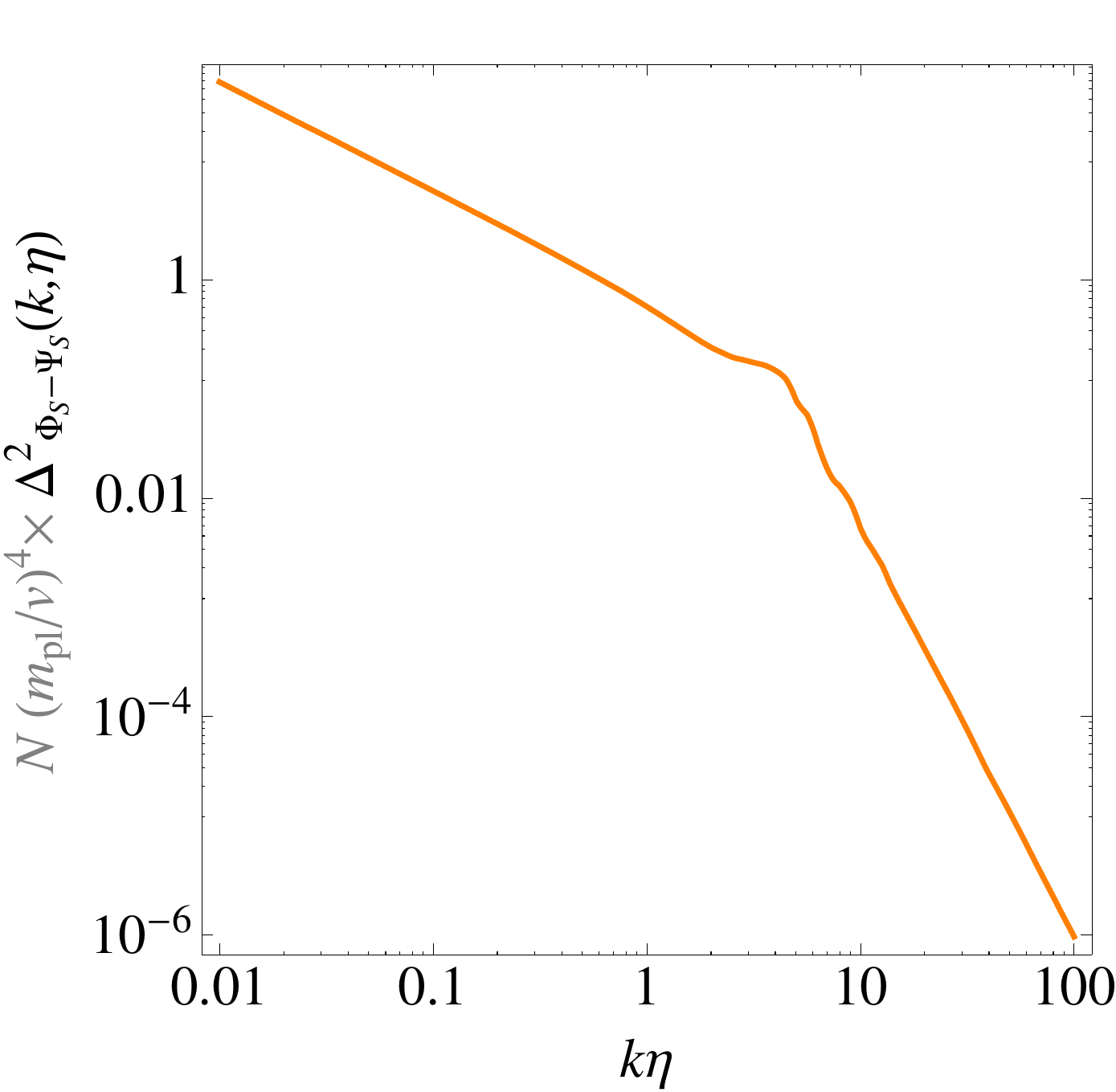} 
   \caption{The dimensionless power spectrum of the difference between the Newtonian and curvature potentials  due to the scalar field only (in a radiation-dominated universe, under the scaling ansatz for the solutions). The spectrum on super horizon scales is $\approx 0.64(k\eta)^{-1}$ whereas on subhorizon scales it is $\approx 30.4(k\eta)^{-4}\ln \left[0.56 (k\eta)\right]$.}   \label{fig:PotentialSpectra2RD}
\end{figure}

In the high-$N$ limit, the central limit theorem and vacuum manifold constraint force the individual field components to be approximately Gaussian distributed\footnote{Non-Gaussian signatures, however, have been computed \cite{Jaffe:1993tt,Gangui:2001fr,2010PhRvD..81l3504F} and used to search for scaling seeds in \textit{Planck} data \cite{Ade:2013xla}.} with zero mean and variance $\propto 1/N$ .  Using Eqs.~(\ref{eq:modevolve}),(\ref{eq:modefunc}),(\ref{eq:psiconvo}),(\ref{eq:phiconvo}) and Eq.(\ref{eq:earlyspec}) along with Wick's theorem allows us to calculate the dimensionless power spectra of the potential fluctuations: 
\begin{widetext}
\begin{align}
\Delta^2_{\Psi_S}(k,\eta)\equiv&\frac{k^{3}}{2\pi^2}P_{\Psi_{S}}(k,\eta)\\
=&\frac{N}{16\pi^2\mpl^4}\frac{1}{k}\int \frac{d^3q}{(2\pi)^3}P_\vp(|\bk+\bq|,\eta_t)P_\vp(q,\eta_t)\nonumber\\
\times&\left[\dot{\phi}_{|\bk+\bq|}\dot{\phi}_q+\bq\cdot(\bq+\bk)\phi_{|\bk+\bq|}{\phi}_q-6\frac{\mH}{k}(\hat{\bk}\cdot\bq)\dot{\phi}_{|\bk+\bq|}\phi_q\right]\nonumber\\
\times&\left[\dot{\phi}_{|\bk+\bq|}\dot{\phi}_q+\bq\cdot(\bq+\bk)\phi_{|\bk+\bq|}{\phi}_q-3\frac{\mH}{k}(\hat{\bk}\cdot\bq)\dot{\phi}_{|\bk+\bq|}\phi_q+3\frac{\mH}{k}(\hat{\bk}\cdot\bq+k){\phi}_{|\bk+\bq|}\dot{\phi}_q\right],\label{eq:psipspec}\\
\Delta^2_{\Phi_S-\Psi_S}(k,\eta)\equiv&\frac{k^{3}}{2\pi^{2}}P_{\Phi_S-\Psi_S}(k,\eta)\nonumber \\=&\frac{9N}{4\pi^2\mpl^4}\frac{1}{k}\int \frac{d^3q}{(2\pi)^3}P_\vp(|\bk+\bq|,\eta_t)P_\vp(q,\eta_t)\phi_{|\bk+\bq|}^2{\phi}_q^2\left[(\hat{\bk}\cdot\bq)^2+\frac{2}{3}(\bk\cdot\bq)-\frac{1}{3}q^2\right]^2.
\label{eq:anisopspec}
\end{align}\end{widetext}
Using Eqs.~(\ref{eq:modefunc}) and (\ref{eq:Aalpha}), these integrals may be numerically evaluated to obtain the curves shown in Figs. \ref{fig:PotentialSpectra1RD}-\ref{fig:PotentialSpectra2RD}. Their amplitudes are proportional to $(v/\mpl)^4/N$, with $v^4$ entering through the initial power spectra of the fields [see Eqns. \eqref{eq:earlyspec} and \eqref{eq:Aalpha}]. The quantity $\Phi_S-\Psi_S$ is proportional to anisotropic stress divided by $k^2$. Note that the dimensionless power spectra of the potentials are functions of $(k\eta)$ only, thus the only scale in the problem is the horizon scale. This is a system that obeys scaling in the sense discussed in the Introduction. Importantly, note that the level of power is preserved on the scale $k\eta\sim 1$ as the network of scaling seeds evolves, a direct consequence of the cubic term in the equations (which in turn arises from the vacuum manifold constraint).

Although there are infrared divergences in the scalar potential $\Phi_S$ and $\Phi_{S}-\Psi_{S}$ power spectra, these metric perturbations are consistent with the fundamental causal requirement that $\left \langle T_{\mu \nu}(\mathbf{x},\eta)T_{\rho \sigma}(\mathbf{y},\eta^{\prime})\right \rangle$ vanish outside the light cone (where $\mathbf{x}$ and $\mathbf{y}$ are two spatial locations). These divergences in metric potentials are a property of many scaling seed models \cite{Durrer:1997te,Durrer:2001cg}.\footnote{The divergences in metric potentials at superhorizon scales can be thought of as  corrections to our local background.} Furthermore, observable quantities of interest such as the gauge invariant radiation density perturbation arising from these seed potentials do not diverge in the infrared. 

Although the fundamental degrees of freedom here are the scalar fields, the stress energy is at lowest-order quadratic in the field amplitude, and so scalars, vectors, and tensor fluctuations in the stress-energy are present \cite{Jaffe:1993tt,Pen:1993nx,Crittenden:1995xf,Durrer:1998rw,Durrer:2001cg,Fenu:2013tea}. They make comparable contributions to CMB anisotropies (in contrast to most inflationary scenarios), and are generally included when imposing limits to scaling seeds and the non-linear $\sigma$ model (such as those stated in the Introduction) from CMB anisotropies. While the constraints from CMB anisotropies include tensor and vector contributions, we will ignore them when calculating the spectral distortions. In this sense our result is likely an underestimate.  

\section{Spectral Distortions}
\label{sec:specd}
Our goal here is to calculate the $\mu$ and $y$ distortions from diffusion damping of acoustic waves sourced by the scaling seeds. Seeds have been considered in previous work, mainly in the context of generating CMB temperature anisotropies \cite{Durrer:2001cg, Perivolaropoulos:1992gy,Durrer:1993db,Pen:1994pe,Durrer:1993db,Crittenden:1995xf,Albrecht:1995bg,Magueijo:1995xj,Hu:1996vq,Turok:1996ud,Magueijo:1996px,Durrer:1997ep,Seljak:1997ii,Durrer:1997rh,Albrecht:1997nt,Durrer:1998rw,Fenu:2013tea,Albrecht:1992sb,Jeong:2004ut,Pogosian:2006hg,Hu:1996yt,Durrer:1997te,Sakellariadou:1999ru,Bevis:2007gh,Ade:2013xla}. Although they are not the dominant source of temperature anisotropies, seeds could still contribute to and perhaps even dominate the spectral distortion signature. 

We begin with an order-of-magnitude estimate of this signature. Recall from Sec. \ref{sec:nonlinearsigma} that the potential generated by the seeds $\Phi_S\sim (v^2/\mpl^2)/\sqrt{N}$. This potential sources acoustic oscillations in the photon fluid at horizon scales with an amplitude $\delta_\gamma\sim \Phi_S$. The spectral distortion amplitude is then determined by energy conservation:  $\mu\sim |\delta_\gamma|^{2}\sim v^{4}/(N m_{\rm pl}^{4})$. Saturating the limits from {\it Planck}, we get $\mu \sim 10^{-10}$. 

We now undertake a more detailed calculation. We first develop the EOMs for linearized perturbations in the distribution of dark matter, baryons, radiation and massless neutrinos in Fourier space. We then numerically solve for the evolution of acoustic modes in the presence of seeds.  This calculation yields a larger value of $\mu\simeq 10^{-9}$, due to the detailed evolution of modes near horizon crossing.

 \subsection{Conservation Equations}
We will continue to work in the conformal Newtonian Gauge, but instead of the usual density perturbations $\delta_n$ in that gauge, we will use
\Beq
\label{eq:udef}
u_n\equiv\delta_n-3(1+w_n)\Psi,
\Eeq
where $w_n\equiv p_n/\rho_n$ is the equation of state for a given species and where $n=\gamma,\nu,{\rm{dm}},b$. These are density perturbations on constant scalar curvature hypersurfaces \cite{Kodama:1985bj}. The following additional definitions reduce clutter in the upcoming equations:
\begin{eqnarray}
\label{eq:defs}
\Lambda &\equiv& \mH/k,\nonumber\\
R_n &\equiv& \rho_n/\rho_{\rm{tot}},\nonumber\\
\Lambda' &=& -\frac{\Lambda^2}{2}\sum_{n}R_n(1+3w_n),\label{eq:Lambda'}\\
A &\equiv&1+\frac{9}{2}\Lambda^2\sum_{n}R_n(1+w_n),\nonumber\\ \nonumber \\
\mathcal{P}_{n} &\equiv& u_n^{\prime}.
\end{eqnarray}
Note that we will be writing the conservation equations in $1$st order form, hence the definition $\mathcal{P}_{n}\equiv u_n'$.  Equation \eqref{eq:Lambda'} above follows from the Friedmann equation. Note that 
\begin{equation}r\equiv k\eta~~\textrm{and}~~^{\prime}\equiv d/d(k \eta). \end{equation}
 The variable $r$ is a natural choice for our independent variable given the scaling behavior of our seed potentials.\footnote{This is not true during the radiation-matter transition, but is valid deep into radiation and matter domination separately.} 

The evolution equations for the different species can be written as follows. See Ref. \cite{Ma:1995ey} for a derivation, though with different notation.\footnote{Schematically, we have taken the equations in Ref.~\cite{Ma:1995ey}, eliminated the density and velocity perturbations $(\delta_n,\theta_n)$ by rewriting the conservation equations in terms of $(u_n,\mathcal{P}_{n})$, used $r=k\eta$ as the time variable, and used the alternative metric convention $(\psi,\phi)\to (\Phi,\Psi)$. The fluid variables in Ref. \cite{Ma:1995ey} are related to ours as follows: $u_n=\delta_n-(1+w_n)\Psi, \theta_n=-k(1+w_n)\mathcal{P}_n$.} 
 \begin{eqnarray}
\mathcal{P}_{\gamma}'&=&\!-\frac{4}{3}\left[\frac{1}{4}u_\gamma\!+\!\Phi\!+\!\Psi\!-\!\sigma_\gamma\right]\!-\!\frac{4}{3\epsilon}\left(\frac{3}{4}\mathcal{P}_{\gamma}\!-\!\mathcal{P}_{b}\right), \label{eq:phocon}\\
 \sigma_\gamma'\!&=&\!-\frac{1}{10}\left[2\mathcal{P}_\gamma\!+\!3F_{\gamma 3}\right]\nonumber\\
& &-\frac{9}{10\epsilon}\sigma_\gamma+\frac{1}{20\epsilon}\left(G_{\gamma 0}+G_{\gamma 1}\right),\label{eq:gammaBoltz2}\\
 F_{\gamma l}'\!&=&\!\frac{1}{2l+1}\left[lF_{\gamma (l-1)}\!-\!(l+1)F_{\gamma (l+1)}\right]\nonumber\\
 & &-\frac{1}{\epsilon}F_{\gamma l}\quad l\ge 3,\label{eq:gammaBoltz}\\
 G_{\gamma l}'\!&=&\!\frac{1}{2l+1}\left[lG_{\gamma (l-1)}\!-\!(l+1)G_{\gamma (l+1)}\right]\nonumber\\
 & &-\frac{1}{\epsilon}\left[G_{\gamma l}-\frac{1}{2}\left(F_{\gamma 2}+G_{\gamma 0}+G_{\gamma 2}\right)\left(\delta_{l0}+\frac{\delta_{l2}}{5}\right)\right]\nonumber \\
\mathcal{P}_{b}'&=&-\Lambda\mathcal{P}_{b}-\Phi+\frac{1}{R\epsilon}\left(\frac{3}{4}\mathcal{P}_{\gamma}-\mathcal{P}_{b}\right),\label{eq:barcon}
\end{eqnarray}
where $\sigma_n$ is the anisotropic stress for the $n^{\rm th}$ species (relevant here for photons and neutrinos), $F_{n2}=2\sigma_n$, $F_{n1}=-\mathcal{P}_{n}$ and $R= (4/3)(\rho_b/\rho_\gamma)$. We have ignored the baryon sound speed since its effects are negligible during the tight-coupling era. The moments $G_{\gamma l}$ capture polarization effects. The tight-coupling expansion parameter $\epsilon$ used above is 
\Beq
\epsilon\equiv(k/n_e\sigma_T a),
\Eeq
and is the ratio of the mean-free-path of the photons $(n_e\sigma_T)^{-1}$ to the physical wavelength of the perturbations $(a/k)$. Here $\sigma_T$ is the Thomson cross section and $n_e$ is the number density of free electrons. The appearance of the last term in Eqs.~\eqref{eq:phocon} and \eqref{eq:barcon} is due to the energy and momentum exchange between these species because of Thomson scattering.\footnote{Note that although the expressions are in Fourier space, we have dropped the $\bf k$ subscript to reduce clutter.} 

The equations for dark matter and neutrinos which are coupled through gravity to all the other species are given by
\begin{eqnarray}
\mathcal{P}_{\rm{dm}}'&=&-\Lambda\mathcal{P}_{\rm{dm}}-\Phi,\\
\mathcal{P}_\nu'&=&-\frac{4}{3}\left[\frac{1}{4}u_\nu+\Phi+\Psi-\sigma_\nu\right],\label{eq:nucon}\\
F_{\nu l}'\!&=&\!\frac{1}{2l+1}\!\!\left[lF_{\nu (l-1)}\!-\!(l+1)\!F_{\nu (l+1)}\right]\!\!\! \quad l\ge \!2\label{eq:nuBoltz}
\end{eqnarray}
Recall that $2\sigma_\nu=F_{\nu 2}$ and $\mathcal{P}_\nu = -F_{\nu1}$.
These equations for photons, baryons, dark matter and neutrinos equations are valid after neutrino decoupling, on all scales as long as the perturbations remain linear. 

\subsection{Tight Coupling and Silk Damping}
Let us focus on the evolution equations for photons and baryons first Eqs. (\ref{eq:phocon})-(\ref{eq:barcon}). In a Hubble time, the co-moving photon diffusion length scale   $k_D^{-1}\sim (Hn_e\sigma_T)^{-1/2}\propto a^{3/2}$. In the last step we assumed radiation domination. This diffusion causes a decay of the acoustic oscillations for $k>k_D$ (Silk Damping). For $k\ll k_D$, the baryons and photons are tightly coupled, making the baryon and photon velocities equal to each other. 
 
Note that for modes that start getting damped in the $\mu$ era, $\epsilon\ll1$. For $\epsilon\ll1$, the equations above have to be handled with some care. In this regime the EOMs for the photons and baryons simplify considerably. In particular the Boltzmann hierarchy for photons can be truncated as follows \cite{Hu:1997hp}:
\Beq
\label{eq:phoTrunc}
&\sigma_\gamma=-\frac{4}{15}{\epsilon}\mathcal{P}_{\gamma}.
\Eeq
This includes the effects of photon polarization. Furthermore, we can eliminate Eq.~(\ref{eq:barcon}), the evolution equation for baryon perturbations, using a tight coupling expansion. Following the clear exposition of Ref. \cite{Tashiro:2012pp}:
\Beq
\label{eq:Pb}
\mathcal{P}_{b}=(3/4)\mathcal{P}_{\gamma}+\epsilon f(r)+\epsilon^2g(r)+\hdots
\Eeq
Using this \textit{ansatz} and keeping only leading order terms in $\epsilon$, we use Eqs. \eqref{eq:phocon}, \eqref{eq:barcon} and \eqref{eq:phoTrunc} to obtain
\Beq
f(r) &= \frac{R}{1+R}\left[\frac{1}{4}u_\gamma-\frac{3}{4}\Lambda\mathcal{P}_{\gamma}+\Psi\right],\\
g(r) & = -\frac{R}{1+R}\left[\frac{(\epsilon f)'}{\epsilon}+\Lambda f-\frac{4}{15}\mathcal{P}_{\gamma}\right].
\Eeq
Using the above $f$ and $g$ in Eq. \eqref{eq:Pb} and Eq. \eqref{eq:phocon} we obtain (at leading order in $\epsilon$):
\begin{eqnarray}
\!\!\!\!\!\mathcal{P}_{b}\!&=&\!\frac{3}{4}\mathcal{P}_{\gamma}+\epsilon \frac{R}{1+R}\!\left[\frac{1}{4}u_\gamma-\frac{3}{4}\Lambda\mathcal{P}_{\gamma}+\Psi\right]\!\!,\label{eq:pi_b}\\
\mathcal{P}_{\gamma}'\!&=&\!-3Rc_s^2\Lambda \mathcal{P}_{\gamma}-c_s^2u_\gamma-\frac{4}{3}(\Phi+3c_s^2\Psi)+\nonumber\\
& &-4\epsilon c_s^2\left\{\frac{4}{15}\mathcal{P}_{\gamma}+Rf'+3R\Lambda f\right\},\label{eq:phocon1}
\end{eqnarray}
where
\Beq
f'&=\frac{\Lambda}{1+R} f+\frac{R}{1+R}\left[\frac{\mathcal{P}_{\gamma}}{4}-\frac{3}{4}\left(\Lambda\mathcal{P}_{\gamma}\right)'+\Psi'\right].
\Eeq These equations are more general than those in Ref. \cite{Hu:1997hp}, following modes starting with their super-horizon evolution on through to horizon crossing, acoustic oscillation, and diffusion damping. We also allow for distinct gravitational potentials. As we will see below, Eq. \eqref{eq:pi_b} allows us to eliminate the equation for $\mathcal{P}_{b}'$. Unlike photons, neutrinos are decoupled (free-streaming), and so the fluid approximation (truncation of the Boltzmann hierarchy after the $l=2$ moment) cannot be applied.

\subsection{Einstein Equations with Seed Potentials}
We need to complete the above system using the Einstein Equations. From the $00$+$0i$, $i\ne j$ and $0i$ Einstein equations we have
\begin{align}
\label{eq:EE0}
\Psi=&\Psi_S+\frac{1}{2m_{\rm pl}^{2}}\frac{a^{2}}{k^{2}}\sum_{n}\left(\delta T^{0}_{0(n)}-3i\frac{\mH}{k} \hat{k}_{j}\delta T^{0}_{j(n)}\right),\ \\
\Phi=&\Psi+\Phi_S-\Psi_S+\frac{3}{2m_{\rm pl}^{2}}\frac{a^{2}}{k^{2}}\sum_{n}\left(\hat{k}_{i}\hat{k}_{j}-\frac{1}{3}\delta_{i}^{j}\right)\delta T_{j(n)}^{i},\nonumber\\
\Psi'=&-\frac{\mH}{k}\Phi+\Psi_{\rm S}^{\prime}+\frac{\mathcal{H}}{k}\Phi_{\rm S}+\frac{1}{2\mpl^2}\frac{a^2}{k^2}\sum_n i\hat{k}_j\delta T^{0}_{j(n)},\nonumber
\end{align}
where $'\,=d/d(k\eta)$, $\Psi_S$ and $\Phi_S$ are the gravitational potentials generated by the scaling seeds and $\sum_{n}$ is over all the species  $(n=\gamma,\nu,{\rm{dm}},b)$. In our notation, these equations can be rewritten as [see Eqs. \eqref{eq:defs}]:
\begin{eqnarray}
\Psi&=&\frac{1}{A}\left[\Psi_S-\frac{3}{2}\Lambda^2\sum_{n}R_n(u_n-3\Lambda \mathcal{P}_{n})\right],\label{eq:EE}\\
\Phi&=&\Psi+\Phi_S-\Psi_S-\frac{9}{2}\Lambda^2\sum_nR_n(1+w_n)\sigma_n,\nonumber\\
\Psi'&=&-\Lambda\Phi+\Psi_{\rm S}^{\prime}+\Lambda\Phi_{\rm S}-\frac{3}{2}\Lambda^2\sum_nR_n\mathcal{P}_{n}.\nonumber
\end{eqnarray}
The $\Psi$ from the $00+0i$ Einstein equation: Eq. \eqref{eq:EE} can be used in Eq. \eqref{eq:pi_b} to express $\mathcal{P}_{b}$ in terms of $u_n,\mathcal{P}_{n\ne b},\Psi_S$ as follows:
\begin{eqnarray}
\mathcal{P}_{b}&=&\frac{3}{4}\mathcal{P}_{\gamma}\label{eq:pibsub}\\
&+&3\epsilon \frac{R}{A} c_s^2\left[\Psi_S-\frac{3}{2}\Lambda^2R_b\left(u_b-\frac{9}{4}\Lambda \mathcal{P}_{\gamma}\right)\right.\nonumber\\
&+&\left.\frac{A}{4}(u_\gamma-3\Lambda \mathcal{P}_{\gamma})-\frac{3}{2}\Lambda^2\sum_{n\ne b}R_i(u_n-3\Lambda \mathcal{P}_{n})\right].\nonumber
\end{eqnarray}
After substituting for $\mathcal{P}_{b}$ using Eq.~(\ref{eq:pibsub}), our system of equations consists of the following: $u_n'=\mathcal{P}_{n}$ ($n=\gamma,\nu,b,{\rm{dm}}$), the $\mathcal{P}_{\gamma}',\mathcal{P}_{\rm{dm}}',\mathcal{P}_\nu'$ Eqs.~(\ref{eq:phocon}), (\ref{eq:barcon}), (\ref{eq:nucon}), \eqref{eq:phoTrunc} for the photon Boltzmann hierarchy, Eq.~\eqref{eq:nuBoltz} for the neutrino Boltzmann hierarchy and the Einstein Equations, Eqs.~\eqref{eq:EE}. These can now be solved once appropriate initial conditions are specified and $\Psi_S$ and $\Phi_S$ are provided. 

The full system of equations to be numerically solved has the form 
\Beq
\label{eq:system}
\mathcal{L}\vec{X}=\vec{\mS},
\Eeq
where $\vec{\mS}$ is vector consisting of linear combinations of $\Phi_S$ and $\Psi_S$, $\vec{X}=\left\{\mathcal{P}_{\gamma},\mathcal{P}_{\rm{dm}},\mathcal{P}_\nu,u_\gamma,u_b,u_{\rm{dm}},u_\nu,\sigma_\nu,F_{\nu(l)}\right\}$, with $l>2$ and $\mathcal{L}=\mathcal{L}[d/dy,y]$ is first order differential operator. Note that $\mathcal{P}_{b}$ is not part of the $\vec{X}$.

\begin{figure}[t] 
\label{fig:ModeEvolve}
   \centering
   \includegraphics[width=3in]{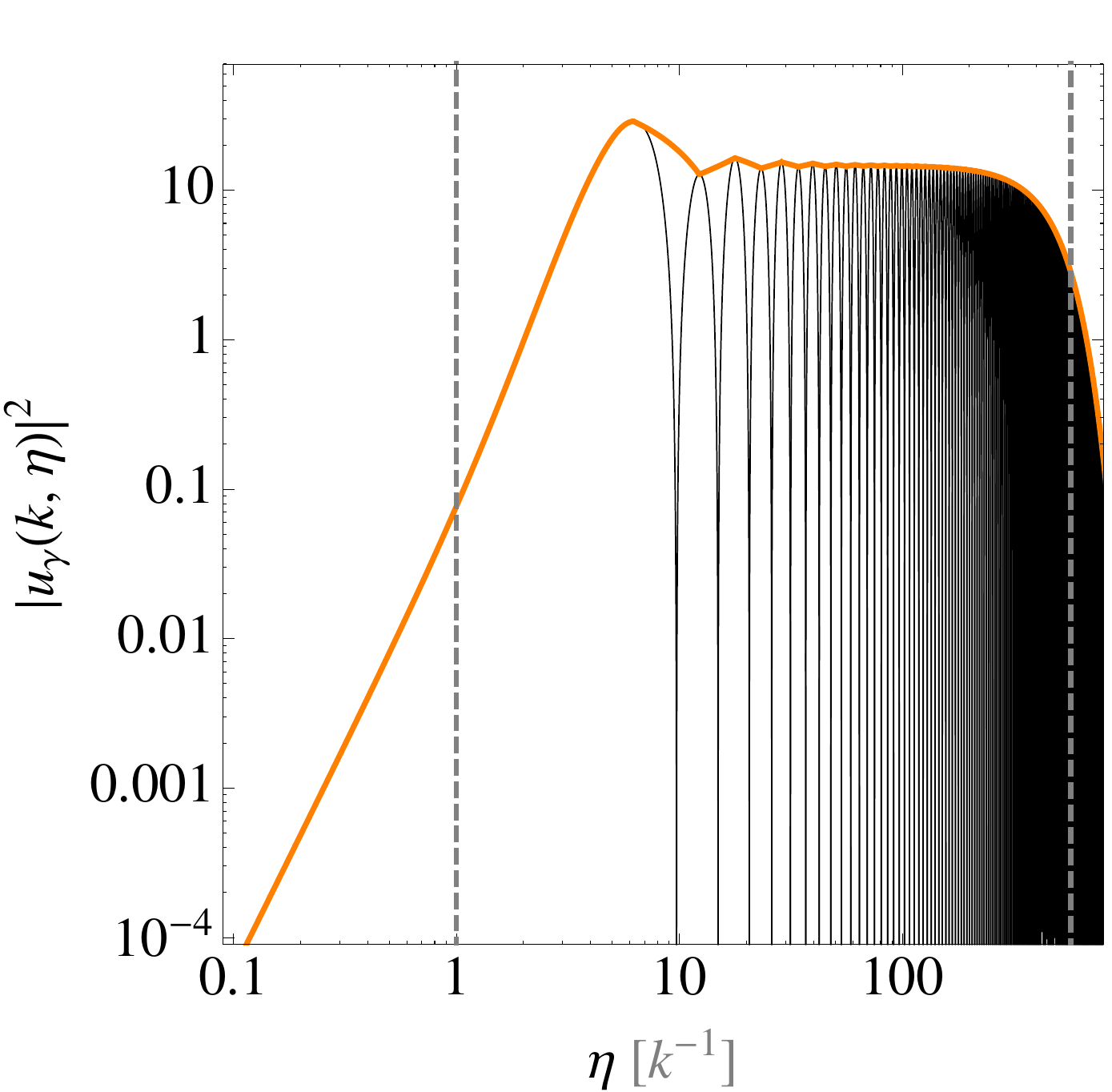} 
   \caption{The evolution of the radiation density perturbations $u_\gamma=\delta_\gamma-4\Psi$ in the presence of seeds, for $k=10^2\, \rm{Mpc}^{-1}$. The black line corresponds to the full evolution of the radiation density perturbation. The orange line follows the full evolution prior to the beginning of acoustic oscillations, and tracks the peak-to-peak envelope of acoustic oscillations once they begin. The dashed line on the left corresponds to horizon entry. Outside the horizon $k\eta\ll1$, $|u_\gamma(k,\eta)|^2\propto \eta^3$. This also implies that the dimensionless power spectrum $\sim k^3|u_\gamma(k,\eta)|^2\propto (k\eta)^3$, corresponding to a scaling, white noise spectrum on superhorizon scales. For $k\eta\gtrsim10$, we see the characteristic acoustic oscillations. For subhorizon scales, $\Psi\ll \delta_\gamma$ even with seed potentials, hence $u_\gamma\approx \delta_\gamma$. Note the continued growth of the photon perturbation for roughly a decade inside the horizon. Finally when the diffusion wavenumber $k_D(\eta)<k$, diffusion takes away the acoustic energy of the mode. The dashed line on the right denotes $k_D(\eta)=k$. }
\end{figure}

\subsection{Solutions}
To compute spectral distortions, we need only the values of photon-related variables, in particular the power spectrum of $u_\gamma$. Distinct fluid components, however, are coupled through gravitational interactions (via Einstein's equations), and so we are forced to solve the entire system simultaneously. 

In the previous section we calculated the power spectra of $\Psi_S$ and $\Phi_S-\Psi_S$. To solve Eqs. \eqref{eq:system}, however, we need the actual mode functions for each Fourier mode of the gravitational potentials. These are not available without numerical field simulations (we do have the mode functions for the scalar field itself, but not the energy momentum tensor or the gravitational potentials). We make the following simplifying \textit{ansatz}, which we justify in an Appendix. We replace $\Psi_S$ and $\Phi_S-\Psi_S$ \cite{Deruelle:1997py},
\Beq
&\Psi_S(\bk,r)\rightarrow \sqrt{\frac{2\pi^2}{k^3}\Delta^2_{\Psi_S}(r)}\,\,e_\bk,\\
&(\Phi_S-\Psi_S)(\bk,r)\rightarrow \sqrt{\frac{2\pi^2}{k^3}\Delta^2_{\Phi_S-\Psi_S}(r)}\,\,e_\bk,
\Eeq
where $e_\bk$ are random variables with $\langle e_\bk e_\bq^*\rangle =\delta(\bk-\bq)$, This is known as the `coherent approximation' \cite{Durrer:2001cg}. 

The two power spectra were calculated in the previous section and are shown in Fig. \ref{fig:PotentialSpectra1RD} and Fig. \ref{fig:PotentialSpectra2RD}. Since we are interested in calculating spectral distortions generated by the seeds, we will set the perturbations in all the components (except the seeds) to be zero initially. Given the linearity of the equations, the solutions we will get will automatically be $f_i(\bk,r) =\sqrt{2\pi^2\Delta^2_{i}(k,r)/k^3}\,e_\bk$ for the component of interest. Here, one should think of $r$ as a time variable. There can be $k$ dependence in the solutions (apart from the $k^{-3/2}$) because damping breaks the scaling nature of the solutions in spite of the scaling behavior of the seed potentials. 

Once the seed potentials, and initial conditions are specified, we can numerically solve Eq. \eqref{eq:system}. For the requisite cosmological parameters we use the current best fit cosmology from \textit{Planck} temperature data (Table 2, last column in Ref. \cite{Ade:2013zuv}). We also use the prescription described in Eq. (51) of Ref. \cite{Ma:1995ey} to cut-off the neutrino hierarchy at $l_{\rm max}=12$. We have made sure that our answers for the spectral distortions are not affected significantly ($<10\%$) by going up to $l_{\rm max}=32$. For comparison, the photon hierarchy was truncated at $l_{\rm max}=2$ because of tight coupling.

The evolution of $u_\gamma(k,r)$ for $k=10^2\,\rm{Mpc}^{-1}$ is shown in Fig. 5. The vertical dashed lines indicate horizon entry and time when the mode starts getting Silk damped [$k_D(\eta)=k$]. On super-horizon scales $u_\gamma \propto \eta^{3/2}$ whereas on sub-horizon scales, we see the characteristic acoustic oscillations as well as exponential damping. We compute this evolution for all modes of interest. In general, acoustic oscillations maintain a fixed amplitude, set essentially by the amplitude at horizon entry, until damping takes over. This is the characteristic behavior of acoustic modes, with and without seed potentials. To understand this note that on subhorizon scales, $u_\gamma =\delta_\gamma-4\Psi\approx \delta_\gamma$. This approximation is valid because gravitational potentials will be suppressed compared to the density perturbations in the dominant component by factors of $(k\eta)^{-2}$ because of Poisson's equation. Hence the solution deep inside the horizon is almost independent of the potential (including the seed potentials), and we are simply seeing the usual acoustic and damping behavior.

%
\begin{figure}[t] 
\label{DiffSpectra}
   \centering
   \includegraphics[width=3in]{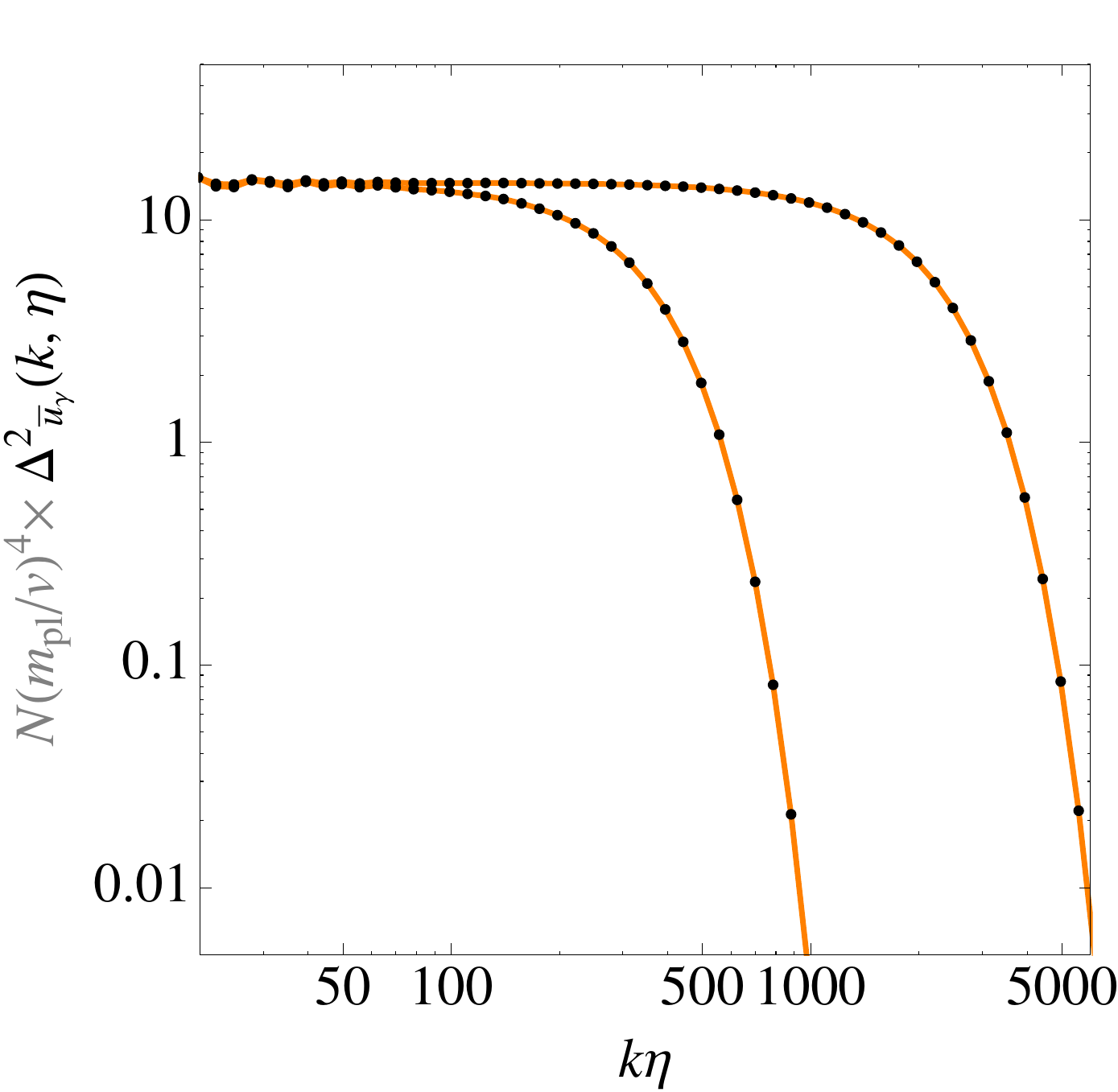} 
   \caption{The envelope of the dimensionless power spectrum $\Delta^2_{\bar{u}_\gamma}(k,\eta)$ of the photon density perturbation evaluated at the beginning and end of the $\mu$ era. Explicitly the top curve corresponds to $\Delta^2_{\bar{u}_\gamma}(k,\eta_f)$ and the bottom curve corresponds to $\Delta^2_{\bar{u}_\gamma}(k,\eta_i)$ where $\eta_i$ and $\eta_f$ correspond to the conformal times at the beginning and end of the $\mu$ era. Black points come from the full evolution code, while the orange curve interpolates between these points. The spectral distortions may be estimated by integrating the difference of the two spectra (with a logarithmic measure). Note that in the text we carry out a more detailed calculation instead of using this integral estimate.}
\end{figure}

While we use a more detailed treatment discussed in the next section, the spectral distortion amplitude can be estimated using the $u_\gamma$ power spectra at the beginning and end of the $\mu$ era \cite{Chluba:2012gq,Khatri:2012rt,Pajer:2012vz}). On sub-horizon scales these spectra show oscillations reflecting the oscillations present in the individual mode solutions. Taking the envelope of these oscillations, the power spectra at the beginning and end of the $\mu$ era are shown in Fig. 6.

With the mode-by-mode solutions of the fluid perturbations generated by the seeds at hand, we are now ready to compute spectral distortions sourced by scaling seeds.

\subsection{$\mu$ Distortion}
During the $\mu$ era, $5\times 10^4\lesssim z\lesssim 2\times 10^6$, double Compton scattering and Bremmstrahlung become inefficient. Acoustic waves damp due to diffusion out of wave fronts on small scales. Neighboring blackbodies are mixed by Thomson scattering, yielding an initial $y$-type distortion. At high $z$, the $y$-type distortion can be partially thermalized and converted into a $\mu$-type distortion by photon energy-changing (single) Compton scattering. 

The damping of acoustic waves heats the plasma, leading to spectral distortions. The fractional heating rate which leads to spectral distortions is given by \cite{Chluba:2012gq}:
\begin{align}
\frac{1}{a^{4}\rho_{\gamma}}\frac{d(a^{4}Q_{\rm ac})}{dz}=&\frac{-4\dot{\tau}\left \langle S_{\rm ac}\right \rangle}{\mathcal{H}(1+z)},\label{eq:heatrate} \\
\left \langle S_{\rm ac} \right \rangle=&\int dk \left [\frac{R^{2}}{1+R}+\frac{16}{15} \right]\frac{k\Delta^{2}_{\mathcal{P}_{\gamma}}(k,\eta)}{16 \dot{\tau}^{2}},
\nonumber\end{align}where $H(z)=H_{0}\sqrt{\Omega_{r}(1+z)^{4}+\Omega_{m}(1+z)^{3}}$, $\Omega_{r}$ and $\Omega_{m}$ are the fractions of the critical density today in radiation and matter, respectively and $\dot{\tau}=n_{e}\sigma_{T}a$ is the conformal-time derivative of the Thomson scattering optical depth. In the above expression, $\Delta^{2}_{\mathcal{P}_{\gamma}}(k,\eta)$ is the time-dependent dimensionless power spectrum of $\mathcal{P}_{\gamma}$. This expression is derived from a more general one in the tight-coupling limit. When waves are diffusion damped, $r=k\eta\gg 1$, and so we justifiably ignore gravitational potentials in the heating formula from Ref. \cite{Chluba:2012gq} and make the approximation $\mathcal{P}_{\gamma}\simeq -F_{1,\gamma}$. We have also ignored higher multipole moments from the Boltzmann hierarchy for photons, which are negligible when $\epsilon \ll 1$.

Assuming that $\mathcal{P}_{\gamma}= \overline{\mathcal{P}}_{\gamma}(k,\eta) f(k,\eta)$ for a smooth envelope $\overline{\mathcal{P}}_{\gamma}(k,\eta)$ and fast oscillatory function $f(k,\eta)$, and averaging over the fast time scale $\int d\eta f^{2}(k,\eta)\simeq 1/2$, the heating rate simplifies to
\begin{eqnarray}
\label{eq:heatingr}
\frac{1}{a^{4}\rho_{\gamma}}\frac{d(a^{4} Q_{\rm ac})}{dz}&\!=&\!\frac{-1}{8\mathcal{H}\eta(1+z)}\!\!\left [\frac{R^{2}}{1+R}+\frac{16}{15} \right] \label{eq:heatratea}\\ &\times &\int_{r_{\textrm{min}}}^{\infty}dr~\epsilon(r,\eta) \Delta^{2}_{\overline{\mathcal{P}}_{\gamma}}(r,\eta),\nonumber
\end{eqnarray}
where $\Delta_{\overline{\mathcal{P}}_{\gamma}}^{2}(r,\eta)$ is the dimensionless power spectrum of $\overline{\mathcal{P}}_{\gamma}$, we have switched to dimensionless wave number $r=k\eta$ as an integration variable and used $\epsilon = k/\dot{\tau}$. In practice we set $r_{\textrm{min}}=10$; below this scale we do not have acoustic oscillations. 

The total $\mu$ distortion generated can be obtained by integrating the heating rate during the $\mu$ era (with a multiplicative factor of $\simeq 1.4$ \cite{Chluba:2012gq}):
\begin{eqnarray}
\mu \simeq 1.4 \int_{z_{\mu,y}}^{z_{\mu}} dz \frac{1}{a^{4}\rho_{\gamma}}\frac{d(a^{4}Q_{\rm ac})}{dz}.\label{eq:simplemu} 
\end{eqnarray} Here $z_\mu\simeq2\times 10^6$ marks the transition from perfect thermalization to $\mu$ distortions while $z_{\mu,y}\simeq 5\times 10^4$ marks the transition into the $y$-distortion epoch.\footnote{It is possible to model the transition between the $\mu$ and $y$-type distortions more carefully using the method described in Ref. \cite{Chluba:2013vsa}. This approach could provide a bridge between the full distortion signal and details about the time-dependence of the energy-injection process \cite{Chluba:2011hw,Khatri:2012tw,Chluba:2013pya}. We find, however, that the heating rate from seed-sourced acoustic waves does not have dramatic features near this transition. We thus defer a computation of this intermediate distortion for future work.} We verified that our numerical implementation of Eqs.~(\ref{eq:heatratea}) and (\ref{eq:simplemu}) reproduces (within $20\%$) the value of $\mu$ in Ref. \cite{Chluba:2012gq} for the adiabatic case with $n_{s}= 1$ and no running.

 We use our actively sourced Boltzmann implementation discussed in the previous subsections to determine $\Delta_{\overline{\mathcal{P}}_{\gamma}}^{2}(r,\eta)$. One of the inputs, the amplitude of the source seed functions, is determined by the combination $(v/\mpl)^2/\sqrt{N}$ (see Figs. \ref{fig:PotentialSpectra1RD} and \ref{fig:PotentialSpectra2RD}). We find that 
\begin{equation}
\mu \simeq 12 \times\frac{1}{N}\left(\frac{v}{\mpl}\right)^{\!4}.
\end{equation} 
This is one of the the main results of our paper. 
Constraints from CMB anisotropies impose the limit $(v/\mpl)^2/\sqrt{N}\lsim 1.3\times 10^{-5}$ \cite{Ade:2013xla}. Saturating this limit, and using our numerically obtained $\Delta_{\overline{\mathcal{P}}_{\gamma}}^{2}(r,\eta)$ in Eq. \eqref{eq:simplemu} we get
\begin{eqnarray}
\mu&\simeq& 2\times 10^{-9}.
\end{eqnarray} 
 
Note that rather than use the familiar approximate damping envelope $e^{-2k^{2}/k_{D}^{2}}$, we have used the full sourced Boltzmann equations to evaluate $\Delta_{\overline{\mathcal{P}}_{\gamma}}^{2}(r,\eta)$. This is done to account for the active sourcing of perturbations after horizon entry in the nonlinear $\sigma$-model. Our more accurate treatment is relevant at times soon after horizon entry, however, deep within the horizon the exponential envelope should still provide a good approximation. This is due to the fact that seed gravitational potentials decay rapidly inside the horizon: $\Phi_S,\Psi_S\sim 1/(k\eta)^2$.

\subsection{ $y$ Distortion}
We now turn to the calculation of the $y$ distortion \cite{Chluba:2012gq}: 
\begin{equation}
y \simeq \frac{1}{4} \int_{0}^{z_{\mu,y}} dz \frac{1}{a^{4}\rho_{\gamma}}\frac{d(a^{4}Q_{\rm ac})}{dz}.\label{eq:simpley} 
\end{equation}
We calculate the $y$ distortion  less accurately than the $\mu$ distortion for both technical and pragmatic reasons. 

On the technical side, note that to evaluate the fractional heating rate, we need to solve the sourced Boltzmann equations. Some of these equations, however, are only valid during the tight coupling era when $\epsilon\ll 1$. This approximation is adequate during the $\mu$ era, however it is violated around decoupling ($z_{\rm dec}\simeq 1090$ \cite{Ade:2013zuv}) which lies within the domain of integration for the $y$ calculation. Moreover, the computation of $Q_{\rm ac}$ after decoupling is complicated by baryon loading, second-order Doppler motion and recombination effects.

We could argue that in the case of passively sourced adiabatic modes, the post decoupling $y$ distortion is several orders of magnitude smaller than the contribution prior to decoupling, and so the integral can be truncated at decoupling:
\begin{equation}
y \simeq \frac{1}{4}\int_{z_{\rm dec}}^{z_{\mu,y}} dz \frac{1}{a^{4}\rho_{\gamma}}\frac{d(a^{4}Q_{\rm ac})}{dz}.\label{eq:simplery} 
\end{equation}
This approximation, however, has not been tested in the case of active sources like $O(N)$ scaling seeds. Additionally, the time dependence of the scalar mode functions changes near matter-radiation equality at $z_{\rm eq}\simeq 3392$ \cite{Ade:2013zuv}, as can be seen from Eqs.~(\ref{eq:modefunc}) and (\ref{eq:besselindex}). In fact, the scaling property of the solutions breaks down during this transition.

On the pragmatic side, the present day $y$ distortion should be dominated by a contribution from reionization of $y \sim 10^{-7}$, unrelated to the primordial signal. As a result the $y$ distortion is not the best probe of primordial physics, though it may someday be possible use cosmological recombination line emission to distinguish primordial $y$ distortions from the signal generated at reionization \cite{Sunyaev:2009qn}.

With these caveats in mind, we would still like to estimate the contribution to the $O(N)$ model $y$-distortion up to decoupling. This, at the very least, requires the evaluation of seed potentials during matter domination as well as a transition in these functions from radiation to matter domination.  As a simple approximation, we impose a switch on the seed potential power spectra $\Delta^{2}_{\Psi_{S}}(k,\eta)$ and $\Delta^{2}_{\Phi_{S}-\Psi_{S}}(k,\eta)$, evaluating them with $\alpha=d\ln a/d\ln \eta=2$ when $z>z_{\rm eq}$ and $\alpha=4$ when $z\leq z_{\rm eq}$. We could also have interpolated between mode functions with some continuous function of $\eta$ as in Refs. \cite{Durrer:1998rw}, but as a first-pass approximation, our method should suffice.\footnote{We have tested that details of `switching' the seed functions at matter-radiation equality do not affect our answer significantly. Even without the switch, that is continuing with seed functions from the radiation dominated era, does not change the $y$ up to decoupling by more than a percent.} 
We find that
\begin{equation}
y\simeq 2.4  \times\frac{1}{N}\left(\frac{v}{\mpl}\right)^{\!4}.
\end{equation}

Saturating the same upper limit from observed CMB anisotropies as used in the $\mu$ case, our estimated value for the $y$ distortion (up to decoupling) from $O(N)$ scaling seeds is
\begin{eqnarray}
y&\simeq&4\times 10^{-10}.\end{eqnarray}
We defer a detailed test of our approximations for future work.
\section{Beyond the $O(N)$ model}
\label{sec:discussion}
So far, we have computed the spectral distortion signature in the large-$N$ limit of the nonlinear $\sigma$-model. It is interesting, however, to consider the spectral distortion signature of a broader class of scaling seed models. As an example, we consider models that are identical to the $O(N)$ case on super horizon scales, but differ from them on subhorizon scales. We parametrize these models by two numbers: $r_*$ and $\gamma$. For $r<r_*$ the seed functions $\Psi_S$ and $\Phi_S-\Psi_S$ are identical to the $O(N)$ case, whereas for $r>r_*$ we allow the slopes of these functions to vary: $\Psi_S,\Phi_S-\Psi_S\sim r^{-\gamma}$. We find that as long as $r_*>10$, the $\mu$ distortion does not change by more than $10\%$ for $-1\le \gamma\le 3$. This shows that for a large class of scaling seed models, as long as the behavior of the seed functions up to a decade in $k$ inside the horizon is similar to the $O(N)$ case, we will have similar $\mu$ distortions. 

We also tried a set of models defined by
\begin{eqnarray}
\Delta^{2}_{\Psi_{S}}(r=k\eta)&=&\frac{\mathcal{A}_{\rm seed}}{r[1+(br)^{2\gamma-1}]},\label{eq:newmodel_a}\\
\Delta^{2}_{\Phi_{S}-\Psi_{S}}(r)&=&c_{\pi}\Delta^{2}_{\Psi_{S}}(r).\label{eq:newmodel_b}
\end{eqnarray}

In the case $\gamma=3.5$ and $c_{\pi}=0$, these models coincide with `Family I' of Refs. \cite{Durrer:2001xu}, designed to reproduce the observed CMB temperature anisotropies (circa the year 2000). 
The power spectra are normalized by the quantity $\mathcal{A}_{\rm seed}$, which is chosen to match the maximum allowed  large-scale normalization of the large-$N$ nonlinear $\sigma$-model considered earlier.\footnote{We note that the maximum allowed $\mathcal{A}_{\rm seed}$ value could in fact be considerably higher, and hence not obey such a tight anisotropy constraint. A class of such models \cite{Turok:1996wa,Hu:1996yt,Albrecht:1997mz,Durrer:2001cg}  were designed to perform more favorably compared to the CMB anisotropy data in absence of inflation. A detailed comparison between these models and present data has not been published.}

To get a sense of how robust our results are to changes in $c_\pi$, $b$ and $\gamma$ we calculated the $\mu$ distortions for a few different values of these parameters. For example, when we fixed $c_\pi=0, b=0.25$ and considered $1\lesssim\gamma \lesssim3.5$, or we fixed $c_\pi=0,\gamma =2$ and considered $0.01\lesssim b\lesssim 0.3$, we still found that $\mu\sim \mathcal{A}_{\rm seed}^2$. For $\gamma=2$ and $b\gtrsim 0.3$, the $\mu$ value starts decreasing rapidly with $b$. This is because as $b$ increases beyond $0.3$, it significantly decreases the power in the seed potentials in the first decade inside the horizon. This is consistent with our analysis of the cutoff scale $r_*$ discussed earlier.
\section{Conclusions}
\label{sec:conclusions}
The cosmic microwave background anisotropies provide a detailed picture of the conditions in the early universe for $k\lsim 10^{-1}\, \textrm{Mpc}^{-1}$. Damping of acoustic modes due to diffusion (and foregrounds) robs us of a chance to get further information on smaller length scales. This very damping, however, leads to distortions of the blackbody spectrum. Thus, spectral distortions allow us to recover some of the lost information on these very small scales: $50\lesssim k\lesssim 10^4\, \textrm{Mpc}^{-1}$. These length scales have never been probed empirically in the linear regime. 

In the standard adiabatic scenario, if we assume an almost scale-invariant initial spectrum of perturbations, the distortions are $\sim 10^{-8}$. This in itself provides an exciting target for future missions. Distortions can be generated in many different ways. One possibility is that components that only contribute sub-dominantly to anisotropies might contribute significantly to the distortion. The magnitude of such a contribution depends on the amplitude and scale dependence of the perturbations. 

In this work, we have explored a scenario where the density perturbations generated by global phase transitions in the early universe damp due to photon diffusion and give rise to spectral distortions of the CMB. These perturbations generated by global phase transitions also influence the CMB anisotropies. When the CMB anisotropy constraints to the $O(N)$ model are saturated, we find that the $\mu$ type spectral distortion is $\mu \simeq 2\times 10^{-9}$. Although we worked with a specific model, the $O(N)$ nonlinear sigma model with $N\gg 1$, we have shown that our result should hold for a much broader class of models to within an order of magnitude. We also estimated the $y$ type signal up to decoupling, and found $y\simeq 4\times 10^{-10}$. 

We made a few simplifying assumptions in our calculation of these spectral distortions. First, we considered equal time correlators, rather than the full unequal time correlators of the seed potentials. We have argued in the Appendix that this could decrease the final answer by a factor of a few at most. It would be useful to check this approximation more carefully by calculating the distortions using the full unequal time correlators. Second, our $y$ estimate can be improved by extending the calculation to the present time by including, for example, effects of free streaming after decoupling. Seed functions that correctly interpolate between the matter and radiation era would also improve the estimate. 

Finally we only considered scalar perturbations. In many defect models and in particular the $\mathcal{O}(N)$ model, the tensor and vector perturbations are comparable to the scalar ones. While tensor and vector contributions were used in determining limits from the CMB anisotropies, we did not include them in the calculation of the spectral distortions. We expect their contribution to be less than the scalar case since these perturbations damp inside the horizon (though not as rapidly as the inflationary case because of the active sourcing). Including them will likely enhance our signal. We will carry out a detailed calculation of vector and tensor contributions to spectral distortions in future work. 

\begin{acknowledgments}
We acknowledge useful conversations with P.~Adshead, D.~Baumann, C. Bonvin, A. Challinor, A.~Guth, M. Hertzberg, W.~Hu, A. Jaffe, M.~Kamionkowski, D.N.~Spergel and A.~Stebbins. We thank Y.~Ali-Ha\"{i}moud and J.~Chluba for useful conversations and a careful reading of this manuscript. MA is supported by a Kavli fellowship at Cambridge. DG is funded at the University of Chicago by a National Science Foundation Astronomy and Astrophysics Postdoctoral Fellowship under Award NO. AST-1302856. DG was also supported at the Institute for Advanced Study by the National Science Foundation (AST-0807044) and NASA (NNX11AF29G). All numerical computations were performed with Mathematica 8.0.
\end{acknowledgments}
\appendix*
\section{Unequal Time Correlators}
\label{sec:Appendix}
In the text we calculated the response of photons, baryons, dark matter and neutrinos to the gravitational potentials generated by the seeds by replacing the $\Psi_S$ and $\Phi_S-\Psi_S$ by the square root of their respective power spectra. Here we estimate the error induced by this approximation.

To reveal the structure of the equations and justify the above simplification, let us ignore all components except photons, and treat them as a perfect fluid with $w_\gamma=c_s^2=1/3$. The conservation and Einstein equations become
\Beq
\label{eq:phoconEins}
&u_\gamma'=\mathcal{P}_{\gamma},\\
&\mathcal{P}_{\gamma}'= -\frac{1}{3}u_\gamma-\frac{4}{3}(\Phi+\Psi),\\
&\Psi=\frac{1}{1+\frac{3}{8r^2}}\left[\Psi_S-\frac{3}{2r^2}\left(u_\gamma-\frac{3}{r}\mathcal{P}_{\gamma}\right)\right],\\
&\Phi=\Psi+\Phi_S-\Psi_S.
\Eeq
Substituting the potentials into the conservation equations we have
\Beq
\label{eq:phoconS}
&u_\gamma'=\mathcal{P}_{\gamma},\\
&\mathcal{P}_{\gamma}'+\left(\frac{12}{r(r^2+6)}\right)\mathcal{P}_{\gamma}+\left(\frac{r^2-6}{3(r^2+6)}\right)u_\gamma=\mS,\\
\Eeq
where
\Beq
\mS=-\frac{4}{3}\left[\frac{2r^2}{r^2+6}\Psi_S+(\Phi_S-\Psi_S)\right].
\Eeq
This equation can be immediately solved to yield
\Beq
u_\gamma=\int_{0}^{r}dv G(r,v)\mS(v),
\Eeq
where
\Beq
G(r,v)&
=\frac{\sqrt{3} v}{r(6+v^2)}\left[(12+vr)\sin\left(\frac{r-v}{\sqrt{3}}\right)\right.\\
&\qquad\left.-6\left(\frac{r-v}{\sqrt{3}}\right)\cos\left(\frac{r-v}{\sqrt{3}}\right)\right].
\Eeq
We have set the homogeneous solutions,which can be interpreted as the inflationary contribution, to $0$.
The dimensionless power spectrum of $u_\gamma$ is then given by
\Beq
\Delta^2_{u_\gamma}(r)=\int_0^r\int_0^r dw dv G(r,w)G(r,v)\Delta^2_\mathcal{S}(w,r),
\Eeq
where $\Delta^2_\mS(w,v)$ is the unequal time, dimensionless power spectrum of $\mS$:
\Beq
\frac{k^3}{2\pi^2}\langle\mathcal{S}_\bk(v)\mathcal{S}^*_\bq(w)\rangle=(2\pi)^3\Delta^2_\mathcal{S}(v,w)\delta(\bk-\bq).
\Eeq
We restored the explicit dependence on the vector aspect of the Fourier momenta above for the sake of clarity. 

In the case of the $O(N)$ model discussed earlier, these unequal time power spectrum can be calculated in terms of fields and their power spectra. From this unequal time correlator, $\Delta^2_{u_\gamma}(y)$ can be calculated using the above Green's functions.

The unequal time correlator is time consuming to evaluate. This source becomes a vector rather than a single function, when dealing with multiple species. In addition, for the system which includes baryons, dark matter and neutrinos, the Green's functions are not available analytically. Significant time and effort is saved by making the following approximation. First we write down the unequal time correlator for the the source function $S$.
\Beq
\Delta^2_\mS(v,w)
&=\frac{16}{9}\left[\left(\frac{2v^2}{v^2+6}\right)\left(\frac{2w^2}{w^2+6}\right)\Delta^2_{\Psi_S}(v,w)\right.\\
&\left.+\left(\frac{2v^2}{v^2+6}\right)\Delta^2_{\Psi_S(\Phi_S-\Psi_S)}(v,w)\right.\\
&\left.+\left(\frac{2w^2}{w^2+6}\right)\Delta^2_{\Psi_S(\Phi_S-\Psi_S)}(v,w)\right.\\
&\left.+\Delta^2_{\Phi_S-\Psi_S}(v,w)\right]\\
\Eeq
The ansatz of replacing $\Psi_S$ and $\Phi_S-\Psi_S$ by square roots of their respective power spectra (as used in the main body of the text) is equivalent to 
\begin{widetext}
\Beq
\Delta^2_S(v,w)\rightarrow &\frac{16}{9}\left[\left(\frac{2v^2}{v^2+6}\right)\sqrt{\Delta^2_{\Psi_S}(v)}\left(\frac{2w^2}{w^2+6}\right)\sqrt{\Delta^2_{\Psi_S}(w)}+\left(\frac{2v^2}{v^2+6}\right)\sqrt{\Delta^2_{\Psi_S}(v)}\sqrt{\Delta^2_{\Phi_S-\Psi_S}(w)}\right.\\
&\left.+\left(\frac{2v^2}{w^2+6}\right)\sqrt{\Delta^2_{\Psi_S}(w)}\sqrt{\Delta^2_{\Phi_S-\Psi_S}(v)}+\sqrt{\Delta^2_{\Phi_S-\Psi_S}(v)}\sqrt{\Delta^2_{\Phi_S-\Psi_S}(v)}\right]\\
\Eeq
\end{widetext}
for this simple one component system. In this system, we have checked that this replacement changes (enhances) the solution $|u_\gamma|^2$ by a factor of $3$. This makes it plausible that our spectral distortion calculation (within the coherent approximation) including all relevant species (photons, dark matter, baryons and neutrinos) will be within a factor of few of the true answer.

Given the fact that the spectral distortions generated by the $O(N)$ model is within an order of magnitude of the one generated by the usual inflationary + \lcdm scenario, and given the possibility that the spectral index for the inflationary case might have running \cite{Ade:2014xna}, it would be worthwhile to check the result using the full unequal time correlator and the eigenfunction method used in Ref. \cite{Fenu:2013tea}.
%

\end{document}